\documentclass[twocolumn,showpacs,preprintnumbers,amssymb,nofootinbib,aps,prd, eqsecnum ]{revtex4-1}
\usepackage{graphicx, booktabs, epsf, epsfig}
\usepackage{bm} 
\usepackage{amsmath}
\usepackage{subfigure}

\def\nn{\nonumber}

\def\be{\begin{equation}}
\def\ee{\end{equation}}
\def\beq{\begin{eqnarray}}
\def\eeq{\end{eqnarray}}
\def \bsp{\begin{split}}
\def \ensp{ \end{split} }

\def\o{\omega}
\def\IL{\relax{\rm I\kern-.18em L}}
\def\nn{\nonumber}
\def\f{\frac}

\begin{document}

 
\title{Slowly rotating regular black holes with a charged thin shell}

\author{Nami Uchikata$^1$}
\email{nami.uchikata@ist.utl.pt}
\author{Shijun Yoshida$^2$}
\email{yoshida@astr.tohoku.ac.jp}
\affiliation{
$^{1}$ Instituto Superior T\'ecnico, Avenida Rovisco Pais, 1049-001 Lisbon, Portugal\\
$^2$Astronomical Institute, Tohoku University, Aramaki-Aoba, Aoba-ku, Sendai
980-8578, Japan}

\date{\today}

\begin{abstract}
We obtain rotating solutions of regular black holes which are constructed of de Sitter spacetime with the axisymmetric stationary perturbation 
within the timelike charged thin shell and the Kerr-Newman geometry with sufficiently small rotation outside the shell. 
To treat the slowly rotating thin shell, we employ the method developed by de la Cruz and Israel. The thin shell is assumed to be composed 
of a dust in the zero-rotation limit and located inside the inner horizon 
of the black hole solution. We expand the perturbation in powers of the rotation parameter of the Kerr-Newman metric up to the second order. 
It is found that with the present  treatment, the stress tensor of the thin shell in general has anisotropic pressure, i.e., the thin shell cannot be 
composed of a dust if the rotational effects are taken into account. However, the thin shell can be composed of a perfect fluid with isotropic pressure 
if the degrees of freedom appearing in the physically acceptable matching of the two distinct spacetimes are suitably used. We numerically 
investigate the rotational effects on the spherically symmetric charged regular black hole obtained by Uchikata, Yoshida and Futamase in detail.  
\end{abstract}


\maketitle
\newpage
\section{Introduction}
The interior of the black hole is still an open problem in gravitational physics.
Although the singularity theorems \cite{pen, haw,he} predict the inevitable formation of the spacetime singularity 
due to the gravitational collapse to the black hole, the theorems are valid only if certain conditions hold within the classical framework.
For example, the strong energy condition is assumed to prove the singularity theorems.
This condition is given by $R_ {\mu \nu} l^{\mu} l^{\nu} \geq 0$, where $R_ {\mu \nu}$ and $l^{\mu}$ are the Ricci tensor and 
a nonspacelike vector, respectively.
In de Sitter spacetime, which is a  vacuum solution of Einstein equations with a positive cosmological constant, $\Lambda >0$, 
the strong energy condition is violated since $R_ {\mu \nu} l^{\mu} l^{\nu} =\Lambda \,  l^{\mu} l_{\mu} \leq 0$ for $l^{\mu} l_{\mu} \leq 0$.
However, the so-called weak energy condition holds in this spacetime. 
Therefore, we could construct black holes without spacetime singularities, i.e., regular black holes or nonsingular black holes, using spacetime with a positive cosmological constant.

In this study, we focus on regular black hole models containing a regular center inside the black hole event horizon \cite{bro}.
Regular black holes of this type basically have causal structures similar to that of the Reissner-Nordstr\" om black hole 
but with regular centers instead of spacetime singularities \cite{bro,an}. Near the center, they behave like the de Sitter spacetime. 
Similarly to the de Sitter spacetime, in these regular black holes, the weak energy condition is satisfied everywhere, 
but the strong energy condition is violated near the center. Thus, these regular black hole models are outside the scope of
the singularity theorems. In other words, the spacetime singularity-free property of these regular black holes is not inconsistent with the singularity theorems. 

The models of the regular black hole we consider can be divided into two classes from the aspect of smoothness of their spacetime.
The models belonging to one class are characterized by a smooth metric, which was first suggested by Bardeen \cite{bar}. In the present  
paper, we call the regular black hole of this class the Bardeen-type regular black hole. 
The solution is typically described by the gravitational field coupled to the nonlinear magnetic field \cite{bro2,ab, ab2, ab3, more}.
A magnetic monopole behaves like the positive cosmological constant so that the spacetime can become regular everywhere 
if there is a magnetic monopole inside the event horizon near the center of the solution. A key ingredient for the solutions of 
this class is the existence of fields acting as an effective positive cosmological constant inside the event horizon. 
This class of regular black hole solutions was generalized to the uncharged cases \cite{dym, dym2, hay}, charged case \cite{dym3} and rotating case \cite{bambi}. 
Formation and evaporation of Bardeen-type regular black holes were considered by Hayward \cite{hay}. 
Quasinormal modes of Bardeen-type regular black holes were investigated in Ref.~\cite{nino}.

The regular black hole model belonging to the other class is characterized by its spacetime composed of two distinct spacetimes: 
one is de Sitter-like spacetime and the other is black hole spacetime,  matched with a thin shell inside the event horizon. 
Models of this class are motivated by the conjecture suggesting that the spacetime curvature has an upper limit of the order of 
the Planck scale \cite{sa, gl, mar}; e.g., the curvature  invariant is restricted by $R_{\mu \nu \alpha \beta} R^{\mu \nu \alpha \beta} \lesssim l_p^{-4}$ 
with  $R_{\mu \nu \alpha \beta}$ and $l_p$ being the Riemann tensor and the Planck length, respectively.
According to this conjecture, the collapsing matter turns into a de Sitter phase when the curvature reaches the upper limit.
For the regular black hole model of this class, as mentioned before, de Sitter spacetime, whose curvature invariant is given by 
$R_{\mu \nu \alpha \beta} R^{\mu \nu \alpha \beta} =24/L^4$ with $L$ being the de Sitter horizon radius, is frequently considered as a regular core instead 
of a spacetime singularity. This conjecture suggests that the de Sitter horizon radius $L$ for this model is of the order of the Planck length \cite{fmm,bal}.

In this study, we focus on the regular black hole of the second class. For the regular black hole solutions of this second class,  the uncharged \cite{fmm, bal,lake} and 
charged spherically symmetric solutions \cite{lemos,uchi} have been investigated. The radial stability of these solutions is studied and it is shown 
that some regular black hole solutions are stable against small radial disturbances \cite{bal, uchi}.  
Although many studies on the regular black hole of the second class have so far focused on spherically symmetric nonrotating cases, 
it is reasonable to generalize  the regular black hole to rotating cases because astrophysical objects usually have angular momentum due to 
its conservation properties.  
In this study, therefore, we will extend previous analyses of spherically symmetric regular black holes to rotating cases, i.e., cases of 
the rotating regular black hole of the second class. 

As a first step, we focus on the slowly rotating case. Rotational effects are then treated as small perturbations 
around spherical solutions. More specifically speaking, we consider the case of $|a| \ll M$, where $a$ and $M$ are the rotation parameter 
and mass of the black hole solution. The rotational effects up to the second order of $|a/M|$ are taken into account in this study 
because we are concerned with rotational effects on the matter properties of the thin shell like deformation of the shell. 
We assume the thin shell to be a timelike hypersurface as done in Ref.~\cite{uchi}. In order for the timelike stationary thin shell to
 exist inside the event horizon, we need the inner horizon besides the event horizon because such a timelike stationary thin shell has to 
 be inside the inner horizon. This implies that a charged black hole has to be considered as the solution outside the shell because 
a slowly rotating Kerr black hole within the accuracy we assume has the event horizon only. 
In this study, therefore, we assume the charged regular black hole obtained in Ref.~\cite{uchi} to be the unperturbed spherically symmetric solution. 
We basically use a method of solution similar to that given by de la Cruz and Israel \cite{dlc2} for matching between the Kerr spacetime and 
the regular  vacuum solution of Einstein equations under the assumption of slow rotation. However, not only the spacetime but also the electromagnetic 
fields have to be considered in the present case because the thin shell is charged.  Outside and inside the shell, we assume a slowly rotating 
Kerr-Newman solution and a stationarily and axisymmetrically perturbed de Sitter solution, respectively. 
The perturbed quantities associated with the thin shell are derived from the junction conditions for the spacetime and electromagnetic field 
(see, e.g., Refs.~\cite{israel,ba,ku}). 
By using analytically obtained results, we show some numerical examples for the slowly rotating charged regular black holes.

The plan of this paper is the following. :
In Sec. II, we give the formalism for obtaining the slowly rotating charged regular black hole and the analytical results derived in the present study. 
In Sec. III, some numerical results for the slowly rotating charged regular black hole are shown. 
A conclusion is given in Sec. IV. Throughout this paper, we use the units of $c = G = 1$, where $c$ and $G$ are the speed of light and 
the gravitational constant, respectively.

\section{Method of solution} 

\subsection{Basic equations for matching of two distinct spacetimes}
As mentioned before, we construct single spacetime in which two different spacetimes   
are matched with a thin shell. For the reader's convenience, here, we concisely describe the formalism developed in Ref.~\cite{israel,ba}, 
which gives a coordinate-independent prescription for matching of two different spacetimes. 
Let $V^{\pm}$ be the four-dimensional spacetimes that have metrics $g_{\alpha \beta} ^{\pm}$ and systems of coordinates ${x^\pm }^{\alpha}$. 
Let $\Sigma$ be a three-dimensional timelike hypersurface described by intrinsic coordinates $y^a$ and located at the boundaries of
 $V^+$ and $V^-$. Thus, $\Sigma$ is given by ${x^\pm }^{\alpha}={x^\pm }^{\alpha}(y^a)$. 
 Here and henceforth, we use greek and roman lower case letters to describe indices of the four-dimensional 
spacetime and  of the three-dimensional hypersurface, respectively. 
Let $n^{\alpha}$ be the unit normal vector to the timelike hypersurface $\Sigma$. Thus, $n^\alpha$ has to satisfy 
\be
n^{\alpha} n_{\alpha} =1, \quad e^{\alpha} _a n_{\alpha}=0,
\ee
where $e^{\alpha} _a$ is the basis vector on $\Sigma$, defined by 
\be
e^{\alpha} _a =\f {\partial x^{\alpha}} {\partial y^a}\,. 
\ee
Here and henceforth, `$\pm$' is sometimes omitted for the sake of brevity.
The induced metric $h_{ab}$ and the extrinsic curvature $K_{ab}$ associated with $\Sigma$ are, respectively,  defined by   
\be
h_{ab} \equiv g _{\alpha \beta} e^{\alpha} _a e ^{\beta} _b\,, \quad K_{ab} \equiv  - n _{\alpha ; \beta} e^{\alpha} _a e ^{\beta} _b.
\ee
Here and henceforth, we denote the covariant differentiation associated with $g_{\alpha \beta}$ and $h_{a b}$  by the semicolon $(;)$ 
and the stroke $(|)$, respectively. So that $V^+$ and $V^-$ are smoothly joined across $\Sigma$, the  first junction condition, 
given by 
\be
 [h_{ab}] =0,
\label{junc1}
\ee
where $[q] \equiv q|^+ - q|^-$, and $q| ^{\pm}$ is the value of the physical quantity $q$ evaluated on $\Sigma$ by taking the limit from 
$V^{\pm}$, has to be fulfilled. Jump in the extrinsic curvature on $\Sigma$ relates to the energy-momentum tensor on $\Sigma$, 
$S_{ab}$, which is explicitly given by 
\be
S_{ab} ={1 \over 8 \pi } \left( [ K_{ab} ] -h_{ab} [K] \right)\,. 
\label{sab}
\ee
When $S_{ab} \ne 0$, we regard $\Sigma$ as a thin shell because there is the matter field on $\Sigma$.  Equation \eqref{sab} is called 
the second junction condition in this paper. 

When there are electromagnetic fields in $V^\pm$, the junction conditions for electromagnetic fields are also required so that 
the electromagnetic fields are smoothly joined across $\Sigma$, as argued in Ref.~\cite{ku}. The junction conditions for the electromagnetic 
field are summarized as follows: 
\be
[F_{ab}] = 0, \quad \mbox{and} \quad  [F_{an}] =4 \pi j _a, 
\label{elj}
\ee
where
\be
 F_{ab} = F_{\alpha \beta} e^{\alpha}_{a} e^{\beta}_b, \quad \mbox{and} \quad  F_{an} = F_{\alpha \beta} e^{\alpha}_{a} n^{\beta}\,. 
\ee
Here, $F_{\alpha \beta}$ stands for the Faraday tensor, given in terms of the vector potential $A_\mu$ by 
$F_{\alpha \beta}  = \partial _ {\alpha} A_ {\beta } - \partial _ {\beta} A_ {\alpha} $, 
and $j_a $ is the current vector on $\Sigma$. The first and  second conditions in Eq.~\eqref{elj}, respectively imply 
that the tangent components of the Faraday tensor to $\Sigma$ have to be continuous through $\Sigma$ and 
that the current on $\Sigma$ is given by the jump of $F_{an}$ on $\Sigma$. These junction conditions are exactly the same as those 
considered in classical electrodynamics (see, e.g., Ref.~\cite{jackson} ). 

\subsection{Two spacetimes matched with a thin shell}
In order to obtain rotating regular black hole solutions, we match two different spacetimes by using a timelike thin shell. 
Inside and outside the thin shell, stationary axisymmetric regular spacetime with a positive cosmological constant $V^-$ and 
the Kerr-Newman spacetime $V^+$, respectively, are assumed in this study. The matching surface $\Sigma$ lies inside 
the inner horizon in $V^+$ and inside the de Sitter horizon in $V^-$ as considered in Ref.~\cite{uchi}. 
The effects of rotation are assumed to be sufficiently 
small that they can be treated as perturbations around a nonrotating spherically symmetric solution considered in Ref.~\cite{uchi}.
Here, the slow rotation means that  the rotation parameter of the black hole is sufficiently small. Thus, it is useful to 
introduce a smallness parameter representing the strength of the rotational effect as $\epsilon^2 \equiv a^2/M^2 \ll 1$, 
where $a$ and $M$ are the rotation parameter and mass of the black hole. It is physically reasonable to expect 
that the rotation parameter of the black hole solution is of the same order as the angular velocities of the thin shell 
and frame dragging of the spacetime inside the thin shell. This expectation is justified when fully consistent solutions 
are obtained by the matching of two distinct spacetimes with a slowly rotating thin shell. In this study, we assume that
rotating regular black hole solutions have equatorial symmetry and time-azimuth reflection symmetry because 
of the symmetry of the Kerr-Newman solution. Therefore, the deformation of the thin 
shell appears as the second-order effects of  $\epsilon$, which correspond to centrifugal effects due to the thin shell's 
spin. We then obtain all the physical quantities of the regular black hole within an accuracy up to the $\epsilon^2$--order. 

Since rotating charged thin shells are considered, as mentioned before, we assume the exterior spacetime of the shell, $V^+$, 
to be a rotating charged black hole solution i.e., the Kerr-Newman solution in the Boyer-Lindquist coordinate and 
expand it up to the $\epsilon^2$ order as follows: 
\be
\begin{split}
ds^2=
      & \displaystyle {\left\{-f_{RN}(\tilde{r}) + \epsilon^2 \left (M^2 \f {Q^2-2M \tilde{r}  } {\tilde{r}^4}\cos^2 \tilde{\theta} +\f{2 \delta M}{ \tilde{r}} \right )\right \} }d\tilde{t}^2 \\
      & +\left \{ \f{1}{f_{RN}(\tilde{r})}+\epsilon^2 \left (\f{M^2 \cos^2 \tilde{\theta}}{\tilde{r}^2 f_{RN}(\tilde{r})} +\f{ 2 \delta M \tilde{r} - M^2}{\tilde{r}^2 f^2_{RN}(\tilde{r} )} \right ) \right \}d\tilde{r}^2 \\
     & +(\tilde{r}^2+\epsilon^2 M^2\cos^2 \tilde{\theta} )d\tilde{\theta} ^2 \\
     & +\displaystyle { \left [ \tilde{r}^2 + \f{\epsilon^2 M^2}{2\tilde{r}^2}\biggl\{ 2\tilde{r}(M+\tilde{r}) -Q^2\right. } \\ 
     & \displaystyle { \quad\quad\quad\quad\quad\quad\quad\quad +(Q^2-2M\tilde{r})\cos2\tilde{\theta} \biggr\} \biggr ]} \sin^2\tilde{\theta}  d \tilde{\phi}^2 \\
     &+ \displaystyle {2 a\left (\f{Q^2-2M\tilde{r}}{\tilde{r}^2}  \sin^2\tilde{\theta} \right )d\tilde{\phi} d\tilde{t}} +O(\epsilon^3) , 
      \label{out}
\end{split}
\ee
where $ f_{RN}(\tilde{r}) = 1-2M/\tilde{r}+Q^2/\tilde{r}^2$ and $Q$ is the charge of the black hole. In this study, 
we assume that the gravitational mass $M$ for the Kerr-Newman solution can change due to rotational effects of the thin shell. 
This change in $M$ is a second-order effect with respect to the spin parameter $a$. 
Then, we explicitly assume that $M \rightarrow M+\epsilon^2 \delta M$ if the thin shell rotates.  
On the other hand, we require charge conservation for the thin shell -- i.e., $Q$ is independent of the spin parameter $a$. This 
assumption will be explored in Sec.~II.E. The vector potential, $A^+_{\mu}$,  is similarly given by 
\be
\tilde{A}^+ = \left(-{Q\over \tilde{r}}+{a^2 Q \cos ^2 \tilde{\theta} \over \tilde{r}^3}\right) d\tilde{t} 
+ {aQ\over \tilde{r}} \sin ^2 \tilde{\theta}d\tilde{\phi}+O(\epsilon^3) \,.
\ee

We assume the interior spacetime of the shell, $V^-$, to be a stationarily and axisymmetrically perturbed de Sitter solution 
given in the so-called static coordinates. The perturbation is too expanded around the de Sitter solution written in the polar 
coordinates $(t,r,\theta,\phi)$ up to the $\epsilon^2$ order as follows: 
\be
\begin{split}
ds^2=
& -f_{dS}(r)\left(1+2 \epsilon^2 h(r,\theta)\right)dt^2 \\
&\displaystyle{ +f_{dS}^{-1}(r)\left(1+ \f{2  \epsilon^2 m(r,\theta)}{r f_{dS}(r)}\right)dr^2} \\
& + r^2\left(1+2\epsilon^2 k(r,\theta)\right) \\ 
& \times \left\{ d\theta^2 +\sin^2 \theta \left(d\phi -\o(r) dt\right)^2 \right\}+O(\epsilon^3),
\label{in}
\end{split}
\ee
where $f_{dS}(r)=1-r^2/L^2$ with $L$ being the de Sitter horizon radius and $\o(r)=O(\epsilon)$. 
The functions appearing in Eq.~(\ref{in}), $\o(r)$, $h(r,\theta)$, $m(r,\theta)$ and $k(r,\theta)$ 
are regular solutions of perturbed Einstein equations coupled to a positive cosmological constant and electromagnetic fields 
generated by the charge of the shell. In order to achieve separation of variables and complete 
matching up to the $\epsilon^2$ order between the two spacetimes described by Eqs.~(\ref{out}) and (\ref{in}), the functions
$h(r,\theta)$, $m(r,\theta)$ and $k(r,\theta)$ are expanded as follows: 
\be
\begin{split}
& h(r,\theta)=h_0(r)+h_2(r)P_2(\cos\theta),\\
& m(r,\theta)=m_0(r)+m_2(r)P_2(\cos\theta),\\
& k(r,\theta)=k_2(r)P_2(\cos\theta).
\end{split}
\ee
where $P_l$ denotes the Legendre polynomial  of degree $l$ (for this treatment of the metric perturbation, see, e.g., 
Refs.~\cite{hartle,chandra,thorne,friedman,rw}). 
The vector potential $A^-_\mu$ is assumed to be 
\be
\begin{split}
\tilde{A}^- &= \left[-{Q\over R}+\epsilon^2 \left\{ B_0(r)+B_2 (r)P_2(\cos \theta) \right\} \right] dt  \\ 
& \quad + a A_3(r) \sin ^2 \theta d\phi +O(\epsilon^3)  ,
\end{split}
\ee
where $R$ denotes the radius of the thin shell in the limit of $\epsilon\rightarrow 0$.
Analytical expressions for the functions appearing in the solution for $V^-$ are derived from perturbed Einstein-Maxwell equations 
with a positive cosmological constant in Appendix A.

\subsection{Equations of the thin shell $\Sigma$}

In order to match the two spacetimes, first of all, we need to specify the shape of the matching layer $\Sigma$. 
The matching layer or thin shell $\Sigma$ is assumed to be given by 
\be
\tilde{r}=R + \epsilon^2 \zeta(\tilde{\theta})+O(\epsilon^3)\,, \quad r = R + \epsilon^2 \xi(\theta) +O(\epsilon^3)\,, 
\ee
in $V^+$ and $V^-$, respectively. Then, we have 
\be
\begin{split}
& n^+ _ {\mu}  =N^+ ( 0, 1, -\epsilon^2 \zeta_{, \tilde{\theta}}, 0 )+O(\epsilon^3) , \\
& n^- _ {\mu} =N^-( 0, 1, -\epsilon^2 \xi_{,  \theta}, 0 )+O(\epsilon^3),
\end{split}
\ee
where $N^\pm$ is determined by the condition of ${n^\pm}^\mu {n^\pm}_\mu = 1$ and the explicit expressions are given by
\be
\begin{split}
& N^+ =f_{RN}^{-{1\over 2}} \left\{ 1-\epsilon^2 \frac{r^2 \left(M^2-2 \delta M r\right)- M^2 f_{RN} \cos^2\theta}{2 r^2 f_{RN}} \right\} \\ 
&\quad\quad\quad\quad\quad\quad+O(\epsilon^3)\,,\\
& N^- =f_{dS}^{-{1\over 2}} \left(1+\epsilon^2 {m(r,\theta) \over r \, f_{dS}} \right) +O(\epsilon^3) \,.
\end{split}
\ee
We choose $y^a=(T,\Theta, \Phi)$ as the intrinsic coordinates of the shell and assume that the shell, $\Sigma$, is given by  
\be
\begin{split}
{x^+}^\mu &\equiv \tilde{x}^{\mu} \,, \\
&= \left (\tilde{t} , \tilde{r} ,\tilde{\theta}, \tilde{\phi} \right) \,,  \\
&= \left (T , R + \epsilon^2 \zeta(\Theta), \Theta, \Phi \right)+O(\epsilon^3) \,, \\
{x^-}^\mu &\equiv x^{\mu} \,, \\
&= \left (t ,r,\theta, \phi \right) \,, \\
&= \left (A T , R + \epsilon ^2\xi(\Theta),  \Theta + \epsilon^2 l^\Theta(\Theta),  \Phi \right)+O(\epsilon^3) \,.
\end{split}
\ee
Here $\tilde{x}^{\mu}$ and $x^{\mu}$ are the four-dimensional coordinates outside and inside the shell, respectively.
The functions $ \xi(\Theta)$ and $\zeta(\Theta)$ can be chosen not to violate the first junction condition \eqref{junc1} 
as discussed later. 
We further assume that 
\be
A=A_0 +\epsilon^2 A_2(\Theta)+O(\epsilon^3)\,.
\ee
The tangent vectors to $\Sigma$ evaluated outside and inside the shell, $\tilde{e} ^{\mu} _a$ and $e ^{\mu} _a$, are then given by  
\be
\begin{split}
\tilde{e}^{\mu}_{T}&= \left (1 , 0 ,0, 0 \right)+O(\epsilon^3)\,, \\
\tilde{e}^{\mu}_{\Theta}&= \left (0 , \epsilon^2 \zeta_{,\Theta} ,1, 0 \right)+O(\epsilon^3)\,, \\
\tilde{e}^{\mu}_{\Phi}&= \left (0 , 0 ,0, 1 \right) +O(\epsilon^3)\,,
\end{split}
\ee
\be
\begin{split}
e^{\mu}_{T}&= \left (A , 0 ,0, 0 \right)+O(\epsilon^3) \,, \\
e^{\mu}_{\Theta}&= \left (\epsilon^2 A_{2,\Theta} , \epsilon^2 \xi_{,\Theta} ,1+\epsilon^2 l^\Theta_{, \Theta}, 0 \right)+O(\epsilon^3)\, , \\
e^{\mu}_{\Phi}&= \left (0 , 0 ,0, 1 \right)+O(\epsilon^3)\, .
\end{split}
\ee
The functions $ \zeta$ and $\xi$ are expanded as 
\be
\begin{split}
& \xi(\Theta)=\xi_0+\xi_2 P_2(\cos\Theta)\,,\\
& \zeta(\Theta)=\zeta_0+\zeta_2 P_2(\cos\Theta)\,.
\end{split}
\ee

\subsection{First junction conditions}

With the assumptions made before, the nonzero components of the induced metrics on $\Sigma$, $h_{ab}^{\pm}$ are given by
\begin{align}
h_{TT}^+ & =  -f_{RN}(R)+ \epsilon^2 \left \{2\zeta \left (\f{Q^2}{R^3} -\f{M}{R^2}\right ) \right. \nn \\
& \quad\quad \left. +M^2 \f{Q^2- 2 M R}{R^4} \cos^2 \Theta+\f{2\delta M }{R} \right\}  +O(\epsilon^3) \,, \\ 
h_{\Theta\Theta}^+ &= R^2 +\epsilon^2 (2 R \zeta + M^2\cos^2 \Theta)  +O(\epsilon^3) \,, \\
h_{\Phi\Phi}^+ & = R^2 \sin^2 \Theta  \nn \\
& + \epsilon^2 \left [ 2 R \zeta +\f{ M^2}{2R^2} \left\{ 2R^2 + (2 MR -Q^2)( 1- \cos2 \Theta) \right\} \right ] \nn\\
&\times \sin^2 \Theta +O(\epsilon^3) \,, \\
h_{T\Phi }^+ & =h_{\Phi T}^+ =a \f{Q^2- 2 M R}{R^2}    \sin^2 \Theta + O(\epsilon^3),
\end{align}
\begin{align}
 h_{TT}^- & = -A_0^2f_{dS}(R) - 2 \epsilon^2 A_0^2\left \{  h  f_{dS}(R) - \f{R\xi }{L^2} \right \} \nn \\ 
 & -2 \epsilon^2 f_{dS}A_0 A_2 +A_0^2 R^2 \o^2 \sin^2 \Theta +O(\epsilon^3) \,,  \\
h_{\Theta\Theta}^- & = R^2 +\epsilon^2 \left( 2 \xi R + 2 k R^2 + 2R^2 l^{\Theta}_{, \Theta}\right )  +O(\epsilon^3) \,, \\
h_{T\Theta}^- & =h_{\Theta T}^- =- \epsilon^2 A_0   f_{dS}(R )A_{2,\Theta} \, T  +O(\epsilon^3) \,, \\
h_{\Phi\Phi}^- & =R^2 \sin^2 \Theta \nn \\ 
& + \epsilon^2 \left( 2 \xi R + 2 k R^2 + 2R^2 \f{\cos \Theta}{\sin \Theta} l^{\Theta}\right )  \sin^2 \Theta  +O(\epsilon^3) \,, \\
h_{T \Phi}^- & =  -\o A_0 R^2  \sin^2 \Theta  +O(\epsilon^3) \,.
\end{align}

The $\epsilon^0$-order condition is obtained from the first junction condition, $[h_{TT}]= 0$, which is given by 
\be
{A_0}^2={f_{RN}\over f_{dS}} \,. 
\ee
The first junction condition $[h_{T \Phi}]= 0$ gives us the $\epsilon$-order condition,  given by 
\be
\begin{split}
\o &=-a \f{Q^2- 2 M R}{A_0 R^4} \,,  \\
&\equiv a \omega_1 \,,
\end{split}
\ee
where we have used the relation $\omega = {\rm const.}$, which is the regular solution for our problem as shown in Appendix A. 
Here, it should be emphasized that the two spacetimes $V^\pm$ can be matched with a slowly rotating thin shell with accuracy 
up to the order $\epsilon$ because the first junction conditions are fulfilled up to the order $\epsilon$, under our assumptions. 
There is no degree of freedom we can specify up to this order. All the physical quantities are fully determined by the unperturbed 
regular black hole solution and the spin parameter $a$.

The $\epsilon^2$-order conditions required by the first junction conditions are summarized 
as follows: 
\begin{align}
& A_2 = {\rm const.}  \,, 
\label{fjc1} \\
& l^{\Theta} =  \f{M^2}{R^2} \left ( \f{1}{2} -\f{Q^2}{2R^2} + \f{M}{R}\right )\sin \Theta \cos \Theta, \label{fjc2} \\
& \xi_0 -\zeta_0 =\f{M^2}{3R} \left (1+\f{M}{R} -\f{Q^2}{2R^2} \right ), \label{con1}  \\
&k_2 (R)R  =  \f{M^2}{3R} \left (-1-\f{4M}{R} +\f{2Q^2}{R^2} \right )+\zeta_2-\xi_2\,, \label{con2} \\
&h_0 (R)= \f{1}{  f_{RN} (R) }\left\{ \f{MR -Q^2 }{R^3} \zeta_0-\f{ \delta M }{R }  \right.\nn \\
& \quad \quad \quad  \left. + M^2\f{(2 M R -Q^2)(R^2 +4 M R -2 Q^2)}{6R^6} \right\} \nn \\ 
&\quad \quad +\f{ R \xi_0 }{  f_{dS} (R)L^2} -\f{A_2}{A_0}, \label{con3} \\
&h_2(R)= \f{M^2(2 M R-Q^2)}{3 R^4} + \f{M R -Q^2}{f_{RN}(R) R^3} \zeta _2 \nn \\
& \quad\quad +\f{R \xi_2}{L^2 f_{dS}(R)}. \label{con4}
\end{align}
Equation (\ref{fjc2}) gives us the explicit form of $l^{\Theta}$, given in terms of the unperturbed quantities only. 
Note that $l^{\Theta} $ is given as the unique equatorially symmetric solution to the first-order ordinary differential 
equation derived from the first junction conditions.  
As shown in Appendix B, the constant $A_2$, given in Eq.~(\ref{fjc1}), does not appear in ${K^-}^a_b$. Thus, the 
basic properties of the thin shell matter are not affected by values of the constant $A_2$. For the sake of 
simplicity,  in this study, we chose the value of $A_2$ to be 
\be
A_2=0 \,. 
\ee
For the $\epsilon^2$-order quantities, then, 
we have the seven undetermined constants, $\delta M$, $\xi_0$, $\xi_2$, $\zeta_0$, $\zeta_2$, $C_4$, and $D_2$. 
Note that $C_4$ and $D_2$ appear in the $l=0$ regular solution ($h_0(R)$) and the $l=2$ regular solutions 
($h_2(R)$, $m_2(R)$ and $k_2(R)$), respectively (see Appendix A). Among these constants, on the other hand, the first 
junction conditions give us the four constraint equations (\ref{con1})--(\ref{con4}). Equations (\ref{con1}) and 
(\ref{con3}) are constraints for $\delta M$, $\xi_0$, $\zeta_0$, and $C_4$. Equations (\ref{con2}) and (\ref{con4}) 
are constraints for $\xi_2$, $\zeta_2$, and $D_2$. 
Therefore, we can freely choose two constants among  $\delta M$, $\xi_0$, $\zeta_0$, and $C_4$, and 
one constant among  $\xi_2$, $\zeta_2$, and $D_2$ not to violate the four constraint equations (\ref{con1})--(\ref{con4}). 
In principle, these three degrees of freedom in the set of the seven constants, $\delta M$, $\xi_0$, $\xi_2$, $\zeta_0$, 
$\zeta_2$, $C_4$, and $D_2$, which completely  determine the structure of the slowly rotating regular black hole considered, 
can be used to specify or simplify the matter stress-energy tensor on $\Sigma$ as argued later.  
In summary, under our assumptions, the two spacetimes $V^\pm$ can be matched within an accuracy up to the order of $\epsilon^2$ 
if any stress-energy tensor is accepted on $\Sigma$.

\subsection{Matching of the electromagnetic fields}
In order to determine the spacetime structure of $V^-$, we have to know the specific form of the electromagnetic fields 
there. From the vector potential $A_\mu^+$, as already given in Sec.~II.B, we may assume the vector potential $A_\mu^-$ 
to be given by  
\be
\begin{split}
\tilde{A}&= \left\{ -{Q\over R}+\epsilon^2 \left( B_0(r)+B_2 (r)P_2(\cos \theta) \right) \right\} dt \\ 
& \quad\quad  + a A_3(r) \sin ^2 \theta \, d\phi +O(\epsilon^3) \, ,
\end{split}
\ee
where the functions $A_3$, $B_0$, and $B_2$ are solutions of the vacuum Maxwell equations, and their explicit forms 
are derived in Appendix A. The nonzero components of the electromagnetic fields tangent to $\Sigma$ are then given by 
\begin{eqnarray}
F_{\Theta\Phi} ^+ &=&-F_{\Phi\Theta} ^+=\frac{2 a Q \sin\Theta \cos\Theta}{R} +O(\epsilon^3) \,,  \\
F_{T\Theta} ^+ &=&- F_{\Theta T} ^+  \nn \\
&=&\frac{\epsilon^2 Q \left(2 M^2+3R \zeta_2\right)\sin\Theta \cos\Theta}{R^3}
+O(\epsilon^3) \,, \\
F_{\Theta\Phi} ^- &=&-F_{\Phi\Theta} ^- \nn \\ 
&=&2 a \sin\Theta \cos\Theta A_3(R)+O(\epsilon^3) \,,  \\
F_{T\Theta} ^- &=&-F_{\Theta T} ^- \nn \\
&=&3 \epsilon^2 A_0  B_2(R) \sin\Theta \cos\Theta
+O(\epsilon^3) \,, 
\end{eqnarray}
If the electromagnetic fields are assumed to be regular in $V^-$, the functions $A_3$ and $B_2$, are given by
\be
 A_3 =C_2\f {r- L\, \mbox{arctanh} (r/L)}{r} \,,
 \label{A3}
\ee
\be
\begin{split}
B_2 & = \f{C_2 L\o_1}{3 M r^3}\left (3 L r +(r^2-3 L^2) x \right ) \\
&+ \f{a_1}{2  r^3} \left ( 2r^3-3 L^2 r+  3  L^3 f_{dS} x\right )\,,\\
\end{split}
\ee
with $C_2$ and $a_1$ being constants. The $\epsilon$-order matching condition, $[F_{\Theta\Phi}]=0$, 
leads 
\be
C_2 =\f { Q} {R-L \, \mbox{arctanh} (R/L)} \, .
\label{c2}
\ee
From the $\epsilon^2$-order matching condition, $[F_{T\Theta}]=0$, we obtain 
\be
\begin{split}
a_1 & = \f {2Q(2 M^2 + 3 \zeta_2 R)}{3 Z } -\f{2 C_2 \o_1 (3 L R +(R^2-3 L^2) X )}{M Z}\,,
\end{split}
\ee
where
\be
X = \mbox{arctanh}\f{R}{L}, \quad Z = 2 R^3 -3 L^2 R +  3 X L^3 f_{dS}(R)\, .
\ee

Since the junction conditions for the electromagnetic fields $[F_{ab}]=0$ are satisfied within an accuracy up to 
the $\epsilon^2$ order, we see that the electromagnetic fields in $V^\pm$ can be matched through $\Sigma$ 
up to the order $\epsilon^2$. From  the second equations in Eq.~(\ref{elj}), we may evaluate the current 
vectors tangent to $\Sigma$ by using jumps of ${{F^{\pm}}^a}_n$ on $\Sigma$. The nonzero components of 
${{F^{\pm}}^a}_n$ are given by 
\be
\begin{split}
& {F^{+T}}_n = \frac{Q}{R^2{f_{RN}}^{1\over 2}} \\
&+ \frac{\epsilon^2 Q}{6
  R^6 \left(f_{RN}\right)^{3/2} }\Biggl[ 6\delta M R^3+2 M^3 R-M^2 \left(Q^2+4 R^2\right) \\ 
&\quad\quad +6 R \left( 3 M R-Q^2-2 R^2\right) \zeta_0  \\ 
&\quad\quad -2 \biggl\{7 M^2 \left\{R (R-2 M)+Q^2\right\} \\ 
&\quad\quad\quad\quad+3 R \left\{ R (2 R-3 M)+Q^2\right\} \zeta_2
\biggr\} P_2\Biggr] +O(\epsilon^3) \,, 
\end{split}
\ee
\be
{F^{+\Phi}}_n =\frac{a Q}{R^4 {f_{RN}}^{1\over 2}}+O(\epsilon^3) \,, 
\ee
\be
\begin{split}
& {F^{-T}}_n = \frac{\epsilon^2}{3 A_0\, {f_{dS}}^{1\over 2}} \Biggl[ 3 B_0^\prime (R)+2 M^2 \omega_1 A_3 '(R) \\ 
&\quad\quad\quad\quad +\left\{ 3 B_2^\prime (R)- 2 M^2 \omega_1 A_3 '(R) \right\} P_2 \Biggr] +O(\epsilon^3) \,, 
\end{split}
\ee
\be
{F^{-\Phi}}_n = -\frac{a{f_{dS}}^{1\over 2} A_3 '(R)}{R^2}+O(\epsilon^3) \,, 
\ee
where $ B_0$ is the regular solution, given by 
\be
\begin{split}
 B_0 & = a_2 + \f{2 \, L \, \o_1 \, x }{3 M r},
\end{split}
\ee
with $a_2$ being a constant. Note that $a_2$ is determined by the continuity condition of 
$A^\pm_a \equiv {e^\pm}^\mu_a A^\pm_\mu$ on $\Sigma$. 
By using these components of ${F^{\pm}}^a_n$, we may know the surface current on $\Sigma$,
 $\displaystyle {j^a={1\over 4\pi}\, [{F^a}_n]}$. 
The current vector $j^a$ satisfies the equation of the charge conservation, i.e., $j^a_{|a}=0$. Thus, we calculate the total 
charge of the thin shell $\Sigma$ by
\be
\tilde{Q}=\int j^{T} \sqrt{-h} d\Theta d\Phi \,, 
\ee
where $h$ denotes the determinant of $h_{ab}$. If we assume the regular solutions for the electromagnetic fields, we have 
\be
\tilde{Q}=Q+O(\epsilon^3)  \,, 
\ee
because 
\be
\begin{split}
& {F^{- T}}_n = \frac{\epsilon^2}{3 A_0\, {f_{dS}}^{1\over 2}} \left\{ 3 B_2^\prime (R)- 2 M^2 \omega_1 A_3 '(R) \right\} P_2+O(\epsilon^3) \,. 
\end{split}
\ee
This fact shows that our assumption that the charge of the black hole does not depend on the spin parameter $a$ is correct within  
the accuracy we consider. 

\subsection{Second junction conditions: Stress-energy tensor of the thin shell}
The second junction condition says that the stress-energy tensor of the thin shell, $S_{ab}$, is given by the jump of
 the extrinsic curvature $K^\pm_{ab}$ across $\Sigma$ (see Eq.~(\ref{sab})).
Nonzero components of the extrinsic curvature ${K^\pm}^a_b$ up to the order $\epsilon$ are given by
\begin{align}
& {K^+}^T_T= \f{Q^2-M R}{R^3\sqrt{f_{RN}}} + O(\epsilon^2) , \\
& {K^+}^{\Theta}_{\Theta}={K^+}^{\Phi}_{\Phi}= - \f{\sqrt{f_{RN}}} {R} + O(\epsilon^2), \\
& {K^+}^{\Phi}_T =a \f{Q^2 -MR }{\sqrt{f_{RN}} R^5} + O(\epsilon^2) ,\\
& {K^+}^T_{\Phi} =a \f{3MR -2Q^2 }{\sqrt{f_{RN}} R^3} \sin^2 \Theta + O(\epsilon^2) , \\
& {K^+} = \f{3MR -Q^2 -2R^2}{\sqrt{f_{RN}} R^3} + O(\epsilon^2),
\end{align} 
\begin{align}
&  {K^-}^T_T= -\f{f_{dS}^{\prime}}{2\sqrt{f_{dS}}} + O(\epsilon^2) , \\
&  {K^-}^{\Theta}_{\Theta}={K^-}^{\Phi}_{\Phi}= -\f{\sqrt{f_{dS}}} {R} + O(\epsilon^2) , \\
&  {K^-}^{\Phi}_T =A_0\left( \f{\sqrt{f_{dS}}}{R} -\f{ f_{dS}^{\prime}}{2 \sqrt{f_{dS}}}\right ) \o + O(\epsilon^2) ,\\
& {K^-} =- \f{4 f_{dS} -R f_{dS}^{\prime}}{2 R \sqrt{f_{dS}}} + O(\epsilon^2)
\end{align} 
The explicit forms of the $\epsilon^2$-order extrinsic curvature are summarized in Appendix B because their expressions are quite 
complicated.  With these expressions, we may obtain explicit forms of the stress-energy tensor of $\Sigma$. 

The matter three-velocity tangent to $\Sigma$, $u^a$, and the total energy density measured by an observer with $u^a$, $\sigma$, 
are, respectively, defined by the timelike eigenvector and the corresponding eigenvalue of $S^a_b$, as follows: 
\be
S^a _b u^b= -\sigma u^a\,, \quad u^au_a=-1 \,. 
\label{sab2}
\ee
Here, $u^\mu=e^\mu_a u^a$ means the four-velocity of the matter associated with the thin shell. The stress tensor of the thin shell, 
$\gamma _{ab}$, measured by an observer with $u^a$ is defined by 
\be
\begin{split}
\gamma _{ab} = {q_a}^c {q_b}^d S_{cd} ,
\end{split}
\ee
where ${q_a}^b$ is the projection tensor associated with $u^a$, defined by ${q_a}^b ={\delta_a}^b+u_a u^b $.
For a perfect fluid, $\gamma_{ab}$ is proportional to $q_{ab}$, i.e., $\gamma _{ab} = p\,q_{ab}$, where $p$ is interpreted as 
the isotropic pressure of the fluid matter.

From the second junction condition, let us first determine the explicit form of $u^a$. 
The $\epsilon^0$-order relations in Eq.~\eqref{sab2} are obviously satisfied by assuming that 
\begin{align}
& u^a= \left({1\over \sqrt{f_{RN}}},0,0\right) +O(\epsilon)\,, \\ 
&\sigma_0=\f{\sqrt{f_{dS}} -\sqrt{f_{RN}}}{4 \pi R} \,, 
\end{align}
where the quantity with the subscript $0$ denotes the quantity in the no-rotation limit. With this $u^a$, we see that 
the matter of the thin shell can be described as a perfect fluid within an accuracy up to the $\epsilon$--order.  
Thus, we have $p=p_0+O(\epsilon^2)$. The isotropic pressure in the limit of $\epsilon \to 0$ is given by 
\begin{align}
p_0=-\f{1}{8 \pi R^2} \left ( \f{M-R}{\sqrt{f_ {RN}}} + R\f{1-2R^2/L^2 }{\sqrt{f_{dS}}} \right ) .
\end{align}
These relations for spherically symmetric solutions may be obtained from the master equations given in Ref.~\cite{uchi} 
if $p_0=0$ is assumed. The solutions given in Ref.~\cite{uchi} can be then used as the unperturbed solutions 
for the present analysis if we consider the case of $p_0=0$ (as argued in the later sections). 
Since ${S^{\Phi}}_{\Theta}=0$, the $\Phi$ component of Eq. \eqref{sab2} leads to
\be
S^{\Phi} _{T} u^{T} + S^{\Phi} _{\Phi} u^{\Phi}= -\sigma u^{\Phi}.
\ee
This equation is satisfied within an accuracy up to the order of $\epsilon$ if we assume 
$u^{\Phi} = \Omega u^{T}$, where $\Omega = O(\epsilon)$.
Therefore we may obtain  
\begin{align}
& \Omega = -\f{a} {R^2} \nn \\
& \times \f{(Q^2 - M R )\sqrt{f_{dS}}+(Q^2 - 2 M R )  \sqrt{f_ {RN}} }{\sqrt{f_{RN}} R^2  -(R^2-3MR + 2Q^2)   \sqrt{f_{dS}}}  \\
& \equiv a \Omega_1 \,. 
\end{align}
Since $S^{\Theta}_{\Phi} =S ^{\Theta}_{T} = O(\epsilon^3)$, we may obtain $u^ {\Theta} = O(\epsilon^3)$. 
Thus, we have 
\be
u^a = u^{T} (1,0,\Omega) +O(\epsilon^3)  \,. 
\ee
The time component of the matter velocity, $u^{T}$, within an accuracy up to the $\epsilon^2$ order may be 
obtained by the normalized condition of $u^a$, 
\be
h_{ab}u^a u^b =-1 \,. 
\ee
The explicit form of $u^{T}$ is then given by 
\beq
&&u^T= \nn \\
&&{1\over \sqrt{f_{RN}}}\Biggl[1+{\epsilon^2\over R^4 f_{RN}}\Biggl\{  R(Q^2-MR)\zeta_0 \nn \\
&& +R^3\delta M + \frac{ M^2}{6} \bigl\{ Q^2-2 M R \nn \\
&&\quad\quad\quad\quad +4 R^2 (Q^2-2MR)\Omega +2 R^6 \Omega_1^2\bigr\}  \nn \\
&& +\biggl\{ R(Q^2-MR)\zeta_2 +\frac{M^2}{3} \bigr\{ Q^2-2 M R   \\
&& \quad +2R^2(2 M R-Q^2)\Omega_1 -R^6\Omega_1^2\bigr\} \biggr\}P_2  \nn 
\Biggr\}\Biggr]  + O(\epsilon^3) \,. 
\eeq

Once $u^a$ within an accuracy up to the $\epsilon^2$ order is obtained, the $\epsilon^2$-order quantities for the thin shell may be 
calculated. The total energy density $\sigma$ of the shell is given by 
\beq
\sigma &=& S_{ab} u^a u^b  \,, \nn \\
&=&\sigma_0+\epsilon^2 \delta \sigma +O(\epsilon^3)\,, 
\eeq
where $\epsilon^2 \delta \sigma$ denotes the $\epsilon^2$-order perturbation of the total energy density $\sigma$. 
The explicit form of $\delta \sigma$ is given in Appendix C.  
The stress tensor of the shell, $\gamma^a_{\>b}$, may be calculated, and its explicit expressions of its nonzero components are given by 
\begin{align}
\gamma^T _{\> T} &= p_0 q^T _{\> T} +O(\epsilon^3) , \quad  \gamma^T _{\>\Phi}  = p_0 q^T _{\>\Phi}+O(\epsilon^3) , \nn  \\
\gamma^{\Phi} _{\>T}  & = p_0 q^{\Phi} _{\>T} +O(\epsilon^3) , 
\label{gam0} 
\end{align}
\begin{align}
\gamma^ +& = p_0 - \f{\epsilon^2 }{8 \pi R^2}\Biggl[  \sqrt{f_{dS} } h_0^{\prime} R^2 + \f {Q^2 - M R} {R^2 \sqrt{f_{RN} }^3  }\delta M  \nn \\
& \quad\quad\quad\quad - \f{L^2 \xi_0 + \left(L^2 -2 R^2\right)m_0 }{L^2\sqrt{f_{dS} }^3 }\nn\\
&  \quad\quad\quad\quad + \f{3M^2 R+ R^3 -M \left(Q^2 + 3 R^2\right)} {R ^3  \sqrt{f_{RN} }^3  } \zeta _0 \nn \\
& +  \f{ M^2} {3R^2 f_{RN} } \Biggl\{   \f{ \left(R-M\right) \left(2 R^2 -2 M R + Q^2\right)} {2  R ^2  \sqrt{f_{RN} } }  \nn \\
&- \f { \left(Q^2 - 2 M R + R^4 \Omega_1\right) \left\{L^2 \left(Q^2 -2 M R\right) + 2 R^6 \Omega_1\right\}} {L ^ 2 R^3  \sqrt{f_{dS} }  } \nn \\
& - \f {R \left(2 M^2 + 3 Q^2 \right) - M \left(Q^2 + 4 R^2\right) } {  \sqrt {f _ {RN }}}  \Omega_1 \nn \\
& -  \f{ 2 R^3 (M R - Q^2) } {\sqrt { f_{RN}}} \Omega_1 ^2  \Biggr\} \Biggr] \nn \\ 
&- \f{ \epsilon^2 P_2 } {8 \pi R^2} \Biggl[ R^2 \sqrt{f_{dS} } \left( h_2 ^{\prime} + k_2 ^{\prime} \right)  \nn \\
& \quad\quad -\f{\left(3R^2 -2 L^2 \right)\xi_2  + \left(L^2 -2 R^2 \right)m_2}{L^2 \sqrt{f_{dS} }^3 }   \nn \\
& \quad\quad + \f{3M^2 R+R^3 -M (Q^2 +3 R^2) }{  R^3\sqrt{f_{RN} }^3 }  \zeta_2   \nn \\
& \quad\quad -\f{M^2}{3 R^2 f_{RN}} \Biggl\{ - \f{ 2 R^3 (MR - Q^2) \Omega_1 ^2 }{\sqrt {f_{RN}}}  \nn \\
& -\f { (Q^2 - 2 M R + R^4 \Omega_1) \{ L^2 (Q^2 -2 M R) + 2 R^6 \Omega_1\}} {L ^ 2 R^3  \sqrt{f_{dS} }  } \nn \\
& - \f {R \left(2 M^2 +3 Q^2\right) - M \left(Q^2 + 4 R^2\right)} {\sqrt {f_{RN}}} \, \Omega_1 \nn \\ 
& + \f{ \left(M-R\right) \left(2 R^2 - 2 M R + Q^2 \right)}{R^2 \sqrt{f_{RN}}} \Biggr\} \Biggr] +O(\epsilon^3) \,,  \label{gp}
\end{align}
\begin{align}
 \gamma^ -&=- \f{\epsilon^2 \sin^2 \Theta}{ 16\pi R^2 }    \Biggl[ \f {M^2} {R^2 \sqrt {f_{RN}} ^3} \biggl\{ M-R    \nn \\
 &+  \Omega_1 \left\{ M \left(Q^2 + 4 R^2\right) - R\left(2 M^2 + 3 Q^2\right) \right\} \nn \\
 & + 2 R^3 \left(Q^2 -M R\right) \Omega_1 ^2  \biggr\} - \f {3 \xi_2} { \sqrt {f_{dS}}}\nn\\
 & - \f{M^2}{ R^5 L^2  f_{RN} \sqrt {f_{dS}} }\left(Q^2 - 2 M R + R^4 \Omega_1\right)  \nn \\
& \times \left\{ L^2 \left(Q^2 - 2 M R \right) + 2 R^6 \Omega _1\right\} \Biggr] +O(\epsilon^3) \,, \label{gn}
\end{align}
where $\gamma ^{\pm} = (\gamma^{\Phi} _{\>\Phi} \pm \gamma^{\Theta} _{\Theta})/2$. The explicit forms of $\gamma^\pm$ can be chosen 
freely as long as the first junction conditions for $h_{ab}$ given in Sec.~II D are not violated. 

\subsection{A perfect-fluid thin shell}
Here and henceforth, we assume that the shell is made of a perfect fluid because a perfect fluid is one of the simplest and most useful models 
for describing astrophysical matter. For a perfect fluid, the stress tensor is, as mentioned before, given by $\gamma_{ab} = p \, q_{ab}$.
Since $u^{\Theta} =O(\epsilon^3)$, as obtained in the last subsection, we obtain 
$\gamma^{\Theta}_{\> \Theta} =p \, \delta ^{\Theta}_{\> \Theta} +O(\epsilon^3)= p \equiv p_0+ \epsilon^2 \delta p$, where $\epsilon^2 \delta p$ denotes 
the $\epsilon^2$-order perturbation of the pressure $p$.  
Thus, the condition that $\gamma^a_{\,b} $ is given by $\gamma^a_{\,b} = p \, q^a_{\,b}$ is equivalent to  the condition 
$ \gamma^{\Phi} _{\>\Phi}= p \, q^{\Phi} _{\>\Phi}$ (see Eq.~(\ref{gam0}), in which we may confirm that all the components 
of $\gamma^a_b$ other than $\gamma^{\Phi} _{\>\Phi}$ are given in the perfect-fluid form).  This condition leads 
\be
\begin{split}
 \gamma^{\Phi} _{\>\Phi} &= p\,q^{\Phi} _{\>\Phi} \\
 &= (p_0 +\epsilon^2 \delta p) (1 +u^{\Phi}u_{\Phi}) + O(\epsilon^3) \,, \\
 & = p_0 + (\epsilon^2 \delta p +p_0 u^{\Phi}u_{\Phi})+ O(\epsilon^3) \,, 
 \end{split}
 \label{p-con0}
\ee
where the facts, $u^{\Phi}=O(\epsilon)$ and  $u_{\Phi}=O(\epsilon)$, have been used. 
Since $\gamma^{\Theta} _{\>\Theta} = p_0 + \epsilon ^2 \delta p+ O(\epsilon^3) $, the condition (\ref{p-con0}) may be reduced to 
the condition for $\gamma^{\Phi} _{\>\Phi}-\gamma^{\Theta} _{\>\Theta}$, 
\be
\gamma^{\Phi} _{\>\Phi}-\gamma^{\Theta} _{\>\Theta} = p_0 u^{\Phi}u_{\Phi} + O(\epsilon^3) .
\ee
This condition and Eq.~(\ref{gn}) give us $\xi_2$ in terms of the unperturbed and first-order perturbation quantities, given by 
\begin{align}
\xi_2 &=  \nn \\
& \f {M^2 }{3 R^2 f_{RN}}  \Biggl[  \f { \sqrt {f_{dS}}} {\sqrt {f _{RN}}} \biggl\{  M-R  + 2R^3\left(Q^2-MR \right) \Omega_1^2 \nn \\
& \quad\quad  + \left(4 M R^2-2 M^2 R+M Q^2-3 Q^2 R \right) \Omega_1 \biggr\}  \nn \\
&  \quad\quad- \left(Q^2-2 M R+R^4 \Omega_1 \right) \nn \\
&  \quad\quad\quad\quad \times \frac{L^2 \left(Q^2-2 M R\right)+2 R^6 \Omega_1}{L^2 R^3}   \label{con5}  \\
& \quad\quad +8 \pi  R^2 \sqrt{f_{dS}} \left(Q^2-2 M R+R^4 \Omega_1 \right)\Omega_1 \, p_0 \Biggr] \,.  \nn
\end{align}
This condition is a third constraint for the three constants $\zeta_2$, $\xi_2$, and $D_2$ besides Eqs.~\eqref{con2} and \eqref{con4}. 
We may therefore determine values of $\zeta_2$, $\xi_2$, and $D_2$ by solving the three coupled linear algebraic equations \eqref{con2}, 
\eqref{con4}, and \eqref{con5}.
For a perfect fluid, the pressure is basically given by a function of the total energy density, i.e., 
$p=p(\sigma)$. The perturbations of $\sigma$ and $p$ are, as seen from Eqs.~\eqref{gp} and \eqref{gn}, expanded by 
\be
\delta\sigma=\delta\sigma_0+\delta\sigma_2 P_2\,, \quad \delta p=\delta p_0+\delta p_2 P_2\,, 
\ee
where $\delta\sigma_0$, $\delta\sigma_2$, $\delta p_0$, and $\delta p_2$ are independent of $\Theta$. 
If the matter is described by a perfect fluid,  $\delta p_0$ and $\delta p_2$ have to relate with $\delta\sigma_0$ and $\delta\sigma_2$ by 
\be
\left( \delta p_0, \delta p_2\right) =\left( {dp \over d\sigma} \right)_0 \left( \delta\sigma_0, \delta\sigma_2 \right) \,. 
\ee
As argued before, the first junction conditions and the assumption of the perfect-fluid shell fully determine values of 
$\zeta_2$, $\xi_2$, and $D_2$.  
This fact directly means that $\delta\sigma_2$ and $\delta p_2$ are also determined. 
Thus, we have 
\be
\left( {dp \over d\sigma} \right)_0 = {\delta p_2 \over \delta\sigma_2} \,, \label{eos}
\ee
unless $\delta\sigma_2=0$. 
This equation gives us information about the equation of state for the thin shell matter around 
the unperturbed state given by $p_0=p(\sigma_0)$. 
Note that in the present study, the two specific spacetimes are first 
matched across a thin shell $\Sigma$, and then the properties of the matter of the thin shell are determined. Thus, the matter 
properties cannot be specified before the matching is done. 
Equation \eqref{eos} gives us a third constraint for 
$\delta M$, $\zeta_0$, $\xi_0$, and $C_4$, given by
\be
\delta p_0  =  {\delta p_2 \over \delta\sigma_2} \,   \delta\sigma_0 \,. \label{eos2}
\ee

To determine the set of the constants $\delta M$, $\zeta_0$, $\xi_0$, and $C_4$, we need one more constraint equation 
for the four constants besides Eqs.~\eqref{con1}, \eqref{con3}, and \eqref{eos2}. Usually, we require 
(i) $\delta \sigma_0=s$, where $s$ is a constant specified appropriately or (ii) the conservation of the total particle number of the shell 
as the spin parameter $a$ increases from $a=0$.  For the requirement (i), $\delta \sigma_0=s$ becomes the fourth constraint equation. 
By solving the four coupled linear algebraic equations \eqref{con1}, 
\eqref{con3}, \eqref{eos2} and $\delta \sigma_0=s$, we may obtain the set of the four constants 
$\delta M$, $\zeta_0$, $\xi_0$, and $C_4$. The simplest example is to set $s=0$. 
Here and henceforth, as unperturbed solutions, we assume the spherically symmetric charged regular black hole solutions obtained by 
Uchikata {\it et} al.~\cite{uchi}, in which the dust shell is considered, i.e., $p_0=0$. 
The total particle number of the thin shell, $N$, is defined by
\be
N = \int n u^T \sqrt{-h}\, d \Theta \, d \Phi \,,
\ee 
where $n$ is the number  surface density of the fluid.  The change in the total particle number, $\delta N$,  can be given by,
\be
\begin{split}
\delta N  = \int (\delta n u^T \sqrt{-h} +n\delta u^T \sqrt{-h} +n u^T\delta \sqrt{- h} )\, d \Theta \, d \Phi \,.
\label{dN}
\end{split}
\ee
From the first law of thermodynamics, given by 
\be
\f{d n}{d \sigma} = \f{n}{\sigma + p} \,,
\ee
and the fact that $p_0=0$, we may obtain $\delta n$ in terms of $ \delta \sigma$, 
\beq
\delta n={n_0 \over\sigma_0}\, \delta \sigma \,. 
\eeq
Thus, the ratio of $\delta N$ to $N$ may be explicitly given by 
\beq
\f{\delta N}{N} &=&\epsilon^2 \left[\f{\delta \sigma_0}{\sigma_0}+ \f{2 \zeta_0}{R}+\f{M^2}{3 R^2} \right. \nonumber \\ 
&&\left. +\f{1+\Omega_1(2 Q^2 - 4 M R + R^4 \Omega_1 )}{3 R^2 f_{RN}} M^2 \right] \, .
\eeq
Note that in Eq.~(\ref{dN}), the terms related to quadrupole perturbations vanish after the angular integration. 
By assuming the total particle number conservation, $\delta N=0$, we may obtain set of solutions  
$\delta M$, $\zeta_0$, $\xi_0$, and $C_4$. The case of  the requirement (ii) is investigated n detail in the next section. 

As argued before, the slowly rotating solutions of the regular
black hole considered in this study are obtained by solving
the two sets of simple coupled linear algebraic equations whose
coefficients are given by elementary functions. In this study,
therefore, no special technique is required for the numerical
calculations, all the numerical procedures are straightforward,
and no difficulty appears in the numerical procedure. While
any numerical programming languages are employable for
the present numerical calculations, the numerical code used
in this study is written in C.
\section{Numerical results}
%
\subsection{Unperturbed solutions: Spherically symmetric charged regular black holes} 
\begin{figure}[t]
\includegraphics {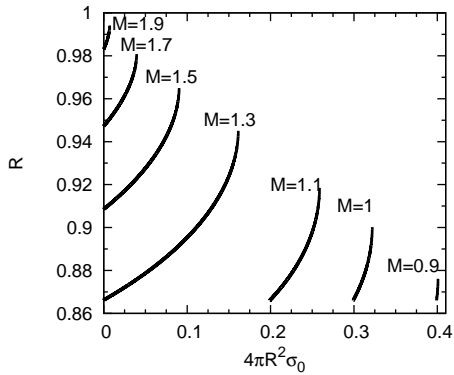}
\caption{Radius of the unperturbed thin shell, $R$, given as a function of the rest mass of the unperturbed thin shell, $4\pi R^2\sigma_0$. 
Each line corresponds to the equilibrium sequence characterized by the same black hole mass, $M$, whose values are given near the corresponding line.}
 \label{fig1}
\end{figure}

\begin{figure}[t]
\includegraphics {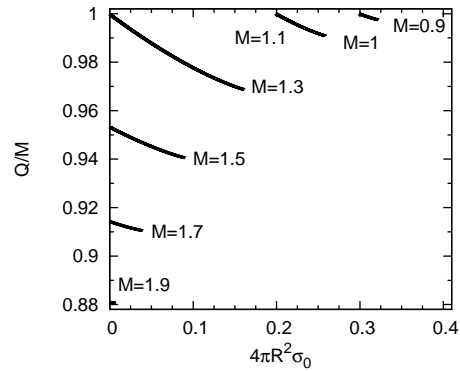}
\caption{Same as Fig.~\ref{fig1}, but for the ratio of the charge to the mass ratio of the black hole, $Q/M$.}
 \label{fig2}
\end{figure}

First, we briefly describe the unperturbed regular black hole considered in this study.   
As discussed in Ref. \cite{uchi}, the equation of motion for a spherically symmetric charged dust thin shell is given by
\be
\sqrt{\dot{R} +1-\f{R^2}{L^2}} -\sqrt{\dot{R} +1-\f{2 M}{R} + \f{Q^2}{R^2}} =4 \pi R \sigma _0,
\label{sph1}
\ee
where $\dot{R} = d R/d \tau$ and $\tau$ is the proper time of the shell. 
Note that in the present paper, $M$ denotes the black hole mass, while it is used for denoting the rest mass of the shell in Ref. \cite{uchi}. 
To investigate motions of the thin shell, it is useful to transform 
Eq.~(\ref{sph1}) to the form
\be
\dot{R} ^2 + V(R) = -1,
\ee
where the effective potential $V$ is defined by 
\be
V(R) = - \left (\f { \f{R^3}{L^2}+\f{Q^2}{R} - 2 M}{8 \pi R^2 \sigma_0} -2 \pi R \sigma_0 \right ) ^2 -\f{R^2}{L^2}\, .
\ee
Then, the stationary solution of the thin shell may be obtained by solving the two algebraic equations, given by 
\be
V(R ) = -1\,, \quad \f{d V(R)}{d R} =0\,,
\ee
simultaneously. Solutions obtained are spherically symmetric charged regular black holes with dust thin shells if the radius of the shell, 
$R$, satisfies the inequalities, $0<R<L$ and $R<r_-$, where $r_-$ denotes the inner-horizon radius of the Reissner-Nordstr\" om black hole. 
For the static solution, the derivative of the effective potential, $dV(R) /dR$, can be rewritten as
\be
\begin{split}
\f{dV(R)} {dR} & = \f {2} { (\sqrt{f_{RN}(R)} -\sqrt{ f_{dS}(R)} )} \\
& \times \left \{ \f {M-R }{R}\sqrt{f_{dS}(R)} + \left ( 1- \f{2 R^2} {L^2} \right )\sqrt {f_{RN}(R)} \right \}\,. 
\end{split}
\ee
Therefore, we may confirm that the condition of  $dV(R)/dR = 0$ is equivalent to the condition of $p_0 = 0$ unless $\sqrt{f_{RN}(R)}=\sqrt{ f_{dS}(R)}$.  
In Figs.~\ref{fig1} and \ref{fig2}, respectively, we show the radius of the shell, $R$, and the ratio of the charge to the gravitational mass 
of the regular black hole, $Q/M$, as functions of the rest mass of the shell, $4\pi R^2\sigma_0$, for the static solutions.  
In these figures, we rescale all the physical quantities by using the length scale of the de Sitter horizon radius $L$, and each line corresponds 
to a sequence of the static solutions of the same black hole mass, $M$, whose values are given near the corresponding lines.
Although stable and unstable solutions are obtained in Ref. \cite{uchi}, in this study, we consider the stable solutions only- i.e., the solutions 
characterized by $d^2V(R) / dR^2 >0$, because unstable solutions are not realized in nature.
%
\subsection{Dipole and quadrupole perturbations}
We numerically evaluate the perturbation quantities for the unperturbed regular black hole solutions, given in Figs.~\ref{fig1} and \ref{fig2}. 
Here and henceforth, all the perturbation quantities are, again, rescaled by using the length scale of the de Sitter horizon radius of the unperturbed solution, $L$. 
As shown in Sec.~II.F, the dipole perturbations, which are of the $\epsilon$-order quantity and associated with $P_1(\cos\theta)$, are uniquely determined by 
the unperturbed quantities only. If the stress tensor describing the thin shell matter is given by the isotropic pressure only, as argued in Sec.~II.G, 
the quadrupole perturbations, which are of the $\epsilon^2$-order quantity  and associated with $P_2(\cos\theta)$, are uniquely determined by the unperturbed 
and the $\epsilon$-order quantities. In other words, we do not have any degree of freedom for obtaining the dipole and quadrupole perturbation 
quantities of the regular black holes with perfect-fluid thin shells. 
To obtain the spherically symmetric perturbation quantities, which are of the $\epsilon^2$-order quantity  and associated with $P_0(\cos\theta)$, on the other hand, 
we have options for selecting what sequences of slowly rotating regular black holes are considered (see the arguments in Sec.~II.G).  
Therefore, we first focus on the results associated with the dipole and quadrupole perturbations. The results associated with 
the spherically symmetric perturbations are then given in the later subsection. 
\begin{figure}[t]
\includegraphics {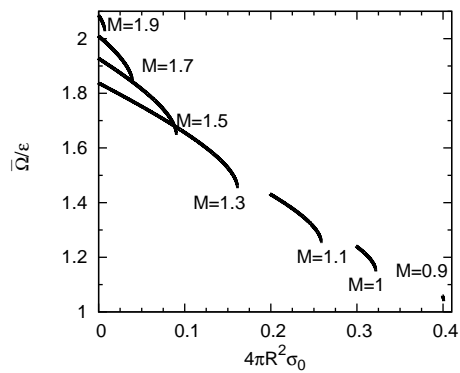}
\caption{Dimensionless angular velocity of the thin shell, $\bar{\Omega}/\epsilon \equiv \Omega/(\epsilon \sqrt{M/R^3})$, given as a function 
of the rest mass of the unperturbed thin shell, $4\pi R^2\sigma_0$. Each line corresponds to the equilibrium sequence characterized by 
the same black hole mass, $M$, whose values are given near the corresponding line.}
\label{fig3}
\end{figure}
\begin{figure}[t]
\includegraphics {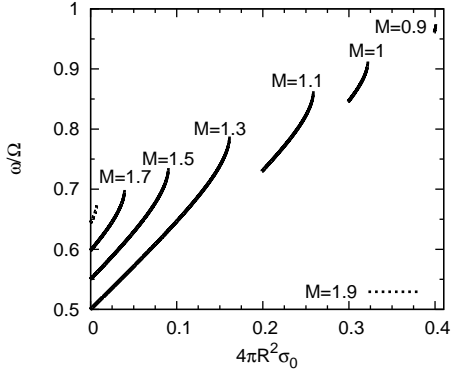}
\caption{Same as Fig.~\ref{fig3}, but for the dimensionless angular velocity of the frame dragging in $V^-$, $\omega/\Omega$.}
\label{fig4}
\end{figure}
\begin{figure}[t]
\includegraphics {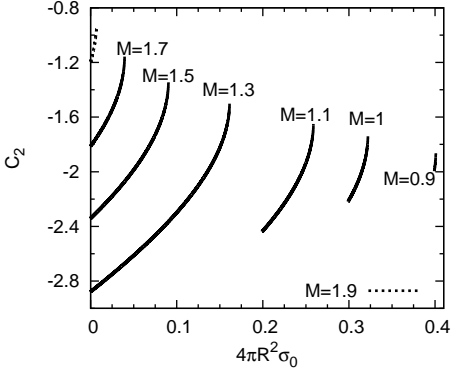}
\caption{Same as Fig.~\ref{fig3},  but for the integration constant $C_2$.}
\label{fig5}
\end{figure}

In Figs. 3 and 4, respectively, show the dimensionless angular velocities of the shell, $\Omega/(\epsilon \sqrt{M/R^3})$, and the frame dragging in $V^-$, 
$\omega/\Omega$, along equilibrium sequences of unperturbed regular black holes given in Figs.~\ref{fig1} and \ref{fig2}. 
When the black hole mass and spin are kept constant,  we observe that the dimensionless angular velocity of the shell, $\Omega$, increases 
as the radius of the shell, $R$, decreases or the charge-to-mass ratio of the black hole, $Q/M$, increases. 
Values of $C_2$, which determine the amplitude of the $\phi$ component of the vector potential, $A^-_\phi$, are given in Fig.~5. 
The constant $C_2$ also determines the amplitude of the magnetic fields, $B^\mu$, at the center of the regular black hole, given by 
\be
\sqrt{B^\mu B_\mu}(r=0)={2\over 3} |a\,C_2|\,. 
\ee
\begin{figure}[t]
\includegraphics {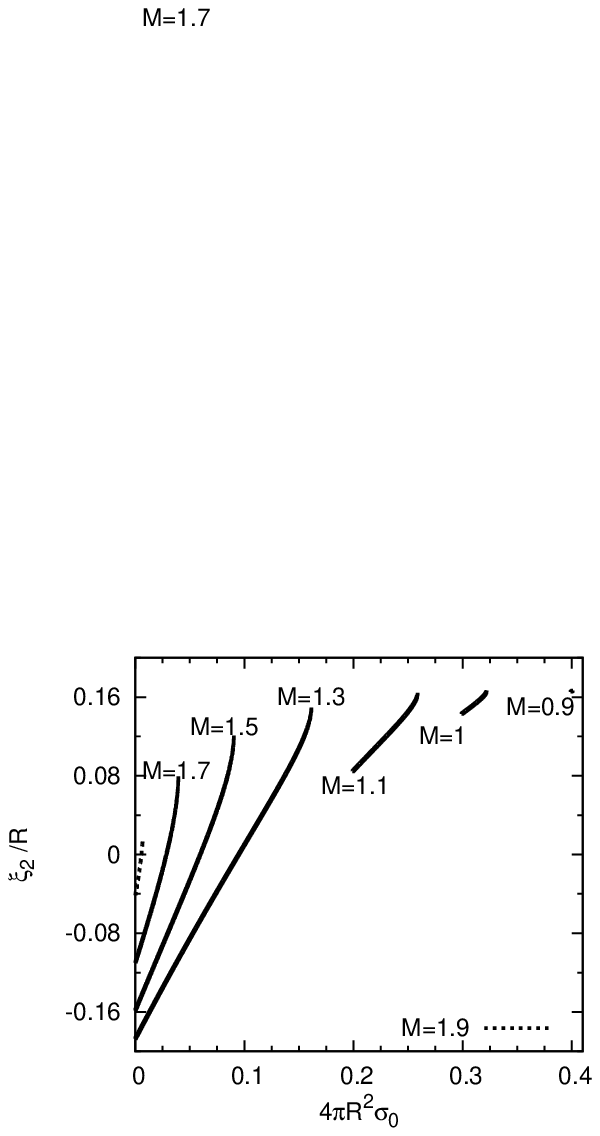}
\caption{Dimensionless quadrupole radial displacement of the thin shell defined in $V^-$, $\xi_2/R$, given as a function 
of the rest mass of the unperturbed thin shell, $4\pi R^2\sigma_0$. Each line corresponds to the equilibrium sequence characterized by 
the same black hole mass, $M$, whose values are given near the corresponding line.}
\label{xi2}
\end{figure}
\begin{figure}[t]
\includegraphics {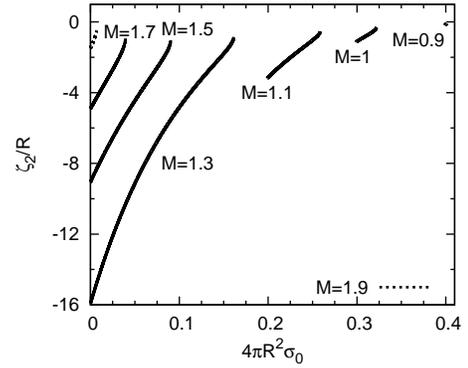}
\caption{Same as Fig.~\ref{xi2}, but for the dimensionless quadrupole radial displacement of the thin shell defined in $V^+$, $\zeta_2/R$.}
\label{zeta2}
\end{figure}
\begin{figure}[t]
\includegraphics {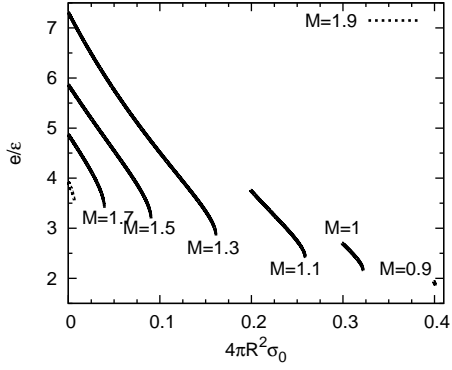}
\caption{Same as Fig.~\ref{xi2}, but for the eccentricity of the thin shell, $e/\epsilon$. }
\label{ecc}
\end{figure}
\begin{figure}[t]
\includegraphics {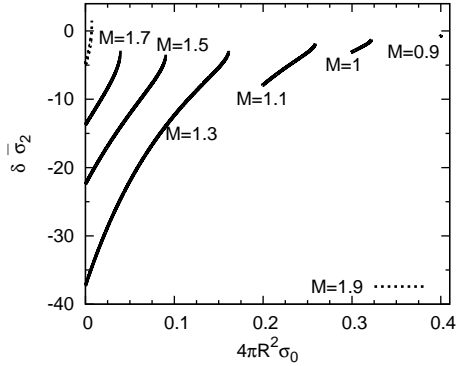}
\caption{Same as Fig.~\ref{xi2}, but for the dimensionless quadrupole energy density perturbation, $\delta\bar{\sigma}_2\equiv\delta\sigma_2/(M/4\pi R^2)$.}
\label{dsig2}
\end{figure}
\begin{figure}[t]
\includegraphics {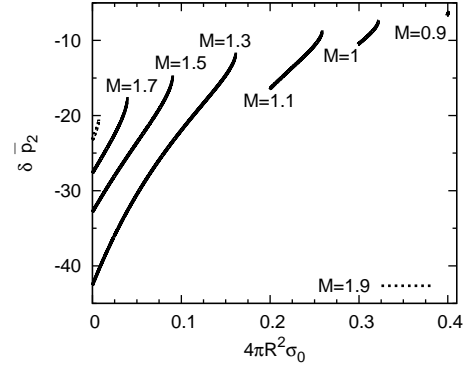}
\caption{Same as Fig.~\ref{xi2}, but for the dimensionless quadrupole pressure perturbation, $\delta\bar{p}_2\equiv\delta p_2/(M/4 \pi R^2)$.}
\label{dp2}
\end{figure}
\begin{figure}[t]
\includegraphics {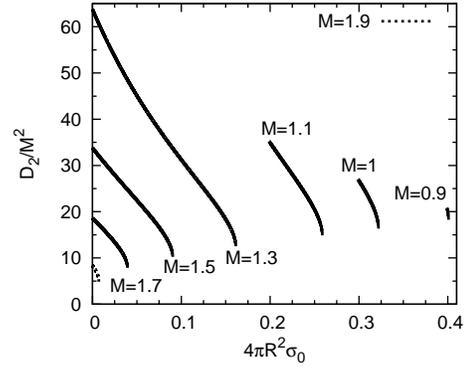}
\caption{Same as Fig.~\ref{xi2}, but for the dimensionless integration constant $D_2/M^2$.}
\label{D2}
\end{figure}

As for the quadrupole perturbation quantities, in Figs. \ref{xi2} through \ref{D2}, the dimensionless radial displacements of the thin shell, $\xi_2/R$ and $\zeta_2/R$, 
the eccentricity of the thin shell, $e/\epsilon$, the dimensionless energy density perturbation of the thin shell, $\delta \sigma_2/(M/4\pi R^2)$, the dimensionless pressure 
perturbation of the thin shell, $\delta p_2/(M/4\pi R^2)$, and the dimensionless integral constant $D_2/M^2$ for the metric perturbations are, respectively, given 
as functions of the rest mass of the unperturbed thin shell, $4\pi R^2\sigma_0$, along the equilibrium sequences given in Figs.~\ref{fig1} and \ref{fig2}. 
Following the standard definition, see, e.g., Refs.~\cite{thorne,friedman}, here, the eccentricity of the thin shell, $e$, is, in terms of the coordinate 
system of $V^-$, given by 
\be
e \equiv \sqrt{ \left(\f{r_e}{r_p}\right)^2-1}= \epsilon \sqrt{- 3\left ( k_2 + \f{\xi_2}{R}\right ) },
\label{deform}
\ee
where $r_e$ and $ r_p$ are, respectively, the effective equatorial and polar radii of the thin shell, given by 
\begin{align}
& r_e = R + \epsilon ^2 \left (\xi_0- \f{\xi_2 +k_2 R}{2} \right ), \\
& r_p = R +  \epsilon ^2 \left (\xi_0+ \xi_2 +k_2 R\right ).
\end{align}
In Figs.~\ref{xi2} and \ref{zeta2}, we observe that $\xi_2$ and $\zeta_2$ show different behaviors; values of $\zeta_2$ are always negative 
but the signs of values of $\xi_2$ depend on unperturbed solutions. However, these behaviors are not directly related to the intrinsic 
physical properties of the thin shell because they are coordinate-dependent quantities defined in the different coordinate systems. Thus, we need quantities defined 
intrinsically in order to see intrinsic physical properties. To see the degree of the deformation of the thin shell, for instance, its eccentricity $e$, 
defined in Eq.~(\ref{deform}), is  a reasonable physical quantity. 
In Fig.~\ref{ecc}, we see that the eccentricity of the thin shell, $e$, is always a real number and all the thin shells are oblately deformed due to rotation 
like standard rotating objects. 
Comparing Fig.~\ref{dsig2} to Fig.~\ref{dp2}, we see that the equilibrium sequence characterized by $M=1.9$ shows some peculiar behaviors. 
In these figures, we observe that near the right end point of the $M=1.9$ sequence, $\delta \sigma_2$ becomes positive, but 
$\delta p_2$ is always negative along the equilibrium sequence. This is an unusual situation because $v_s^2 <0$ if we naively assume the sound 
speed $v_s$ to be given by $v_s^2 =\delta p/ \delta\sigma$.  Although this definition for the sound speed is reasonable, 
we cannot know how to define the actual sound speed of the thin shell matter at the level of the present approximation because we consider 
stationary states only.  Thus, the appearance of this peculiar behavior does not immediately mean a flaw in our treatment. 
Aside from the problem with the sound speed, the present scheme for obtaining the slowly rotating regular black hole with the perfect-fluid thin shell becomes 
unavailable for the unperturbed solution having $\delta\sigma_2=0$ as discussed in Sec.~II.G. Thus, the unperturbed solutions having 
$\delta\sigma_2 \gtrsim 0$ are not considered in the later discussions.  
For other sequences considered in this study, that peculiar behavior with $\delta \sigma_2$ is not observed and values of  $\delta \sigma_2$ 
and $\delta p_2$ are always negative as shown in Figs.~\ref{dsig2} and \ref{dp2}. 
The integral constant $D_2$, given in Fig.~\ref{D2}, basically determines the amplitude of the metric perturbations in $V^-$ 
(for details, see Appendix A).
As shown in this figure, the maximum value of $D_2/M^2$ for unperturbed solutions considered in this study is given by $D_2 /M^2\sim 65$ for 
the unperturbed solution characterized by $M=1.3$ and $\sigma_0 \sim 0$, which might give a slightly stronger restriction for  suitable 
values of $\epsilon$ due to its slightly large maximum value of $D_2/M^2$.

\subsection{Spherically symmetric perturbations for the equilibrium sequences of slowly rotating charged regular black holes characterized 
by the fixed total particle number of the shell} 
%
\begin{figure}[t]
\includegraphics {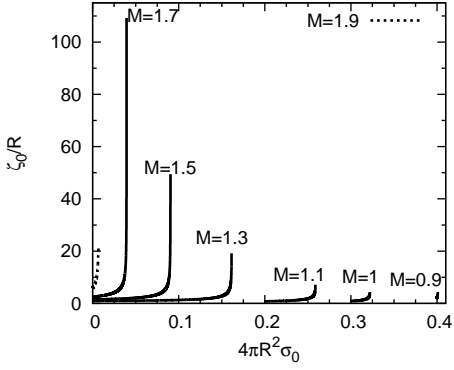}
\caption{Dimensionless spherically symmetric radial displacement of the thin shell defined in $V^+$, $\zeta_0/R$, given as functions 
of the rest mass of the unperturbed thin shell, $4\pi R^2\sigma_0$. Each line corresponds to the equilibrium sequence characterized by 
the same black hole mass, $M$, whose values are given near the corresponding line.}
\label{zeta0b}
\end{figure}
\begin{figure}[t]
\includegraphics {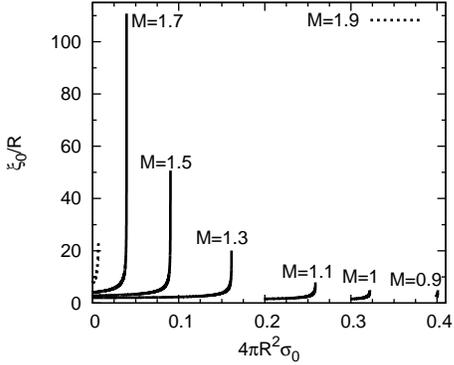}
\caption{Same as Fig.~\ref{zeta0b}, but for the dimensionless spherically symmetric radial displacement of the thin shell defined in $V^-$, $\xi_0/R$.}
\label{xi0b}
\end{figure}
\begin{figure}[t]
\includegraphics {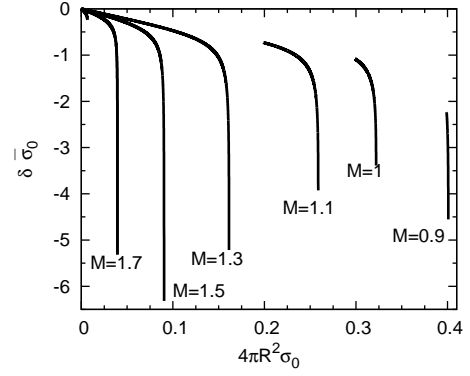}
\caption{Same as Fig.~\ref{zeta0b}, but for the dimensionless spherically symmetric energy density perturbation, 
$\delta\bar{\sigma}_0\equiv \delta\sigma_0/(M/4\pi R^2)$. The short unlabeled line at the left top corresponds to the result
for the M=1.9 sequence.}
\label{dsig0}
\end{figure}
\begin{figure}[t]
\includegraphics {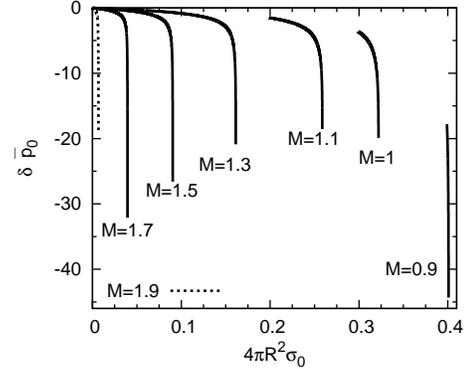}
\caption{Same as Fig.~\ref{zeta0b}, but for the dimensionless spherically symmetric pressure perturbation, $\delta\bar{p}_0\equiv\delta p_0/(M/4 \pi R^2)$.}
\label{dp0}
\end{figure}
\begin{figure}[t]
\includegraphics {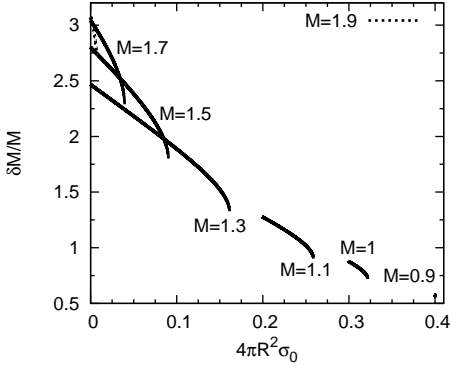}
\caption{Same as Fig.~\ref{zeta0b}, but for the relative changes in the gravitational mass of the back hole, $\delta M/M$.}
\label{dMb}
\end{figure}
\begin{figure}[t]
\includegraphics {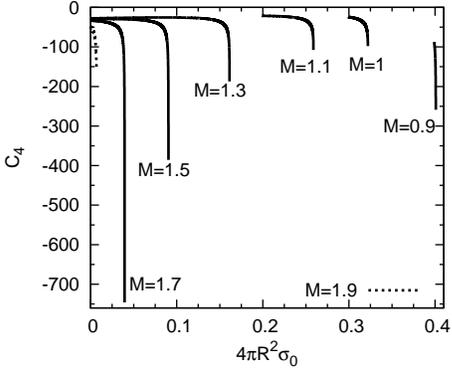}
\caption{Same as Fig.~\ref{zeta0b}, but for the integration constant $C_4$ appearing in $h_0$.}
\label{C4b}
\end{figure}
In order to obtain spherically symmetric perturbations due to rotation, as mentioned before, we assume that the total particle number or 
total rest mass of the charged thin shell is conserved as its rotation increases from zero, i.e., the condition $\delta N/N=0$. 
We then obtain $\delta M$, $\zeta_0$, $\xi_0$ and $C_4$ for the equilibrium sequences given in Figs.~\ref{fig1} and \ref{fig2}.
For the $M=1.9$ equilibrium sequence, as mentioned before, $(dp/d\sigma)_0$ cannot be defined for the unperturbed solution with 
$4 \pi R^2 \sigma_0 \sim 0.007$ because $\delta \sigma_2 $ vanishes. As a result, we cannot obtain the spherically symmetric perturbation 
for the unperturbed solutions having $4 \pi R^2 \sigma_0 \sim 0.007$. For  the $M=1.9$ equilibrium sequence, therefore, we only give 
results for the unperturbed solutions with $\delta \sigma_2 \lesssim  -0.04$ in this paper. 
The values of $\zeta_0/R$, $\xi_0/R$, $\delta \sigma_0/(M/4\pi R^2)$, $\delta p_0/(M/4\pi R^2)$, $\delta M/M$, and $C_4$ are given as functions of 
the rest mass of the unperturbed thin shell, $4\pi R^2 \sigma_0$, in Figs. \ref{zeta0b} through \ref{C4b}, respectively. 
As shown in Figs.~\ref{zeta0b} and \ref{xi0b}, behaviors of $\xi_0$ are similar to those of $\zeta_0$. This is expected by Eq.~(\ref{con1}), given 
by the first junction condition.  
Since the Schwarzschild-like coordinate system is used in $V^-$, we may evaluate the mean radius of the thin shell deformed spheroidally, $\bar{r}$, 
by $\bar{r}=R(1+\epsilon^2\xi_0/R)$. As shown in Fig.~\ref{xi0b}, $\xi_0$'s always take positive values for the equilibrium sequences characterized 
by $\delta N/N=0$. This is the usual situation because the centrifugal force tends to increase the equatorial radius of the rotating shell and its mean 
radius therefore increases if the rest mass of the shell is kept constant. 
In Figs.~\ref{dsig0} and \ref{dp0}, we observe that values of $\delta \sigma_0$ and $\delta p_0$ are always negative and 
approach zero in the $\sigma _0 \to 0$ limit along the equilibrium sequences characterized by a fixed gravitational mass $M$ as long as we may 
take the $\sigma _0 \to 0$ limit. 
As shown in Fig.~\ref{dMb}, the relative changes in the gravitational mass of the black hole, $\delta M/M$ are always positive for the unperturbed 
solutions considered in this study.  
This is because the rotational energy is added in the thin shell but the rest mass and charge of the thin shell is conserved. 
The integration constant $C_4$, given in Fig. \ref{C4b}, basically gives a value of the metric perturbation $h_0$ at the center of the regular black hole. 
Similar to the values of $D_2/M^2$, slightly large values of $C_4$ might give a slightly stronger restriction for  suitable 
values of $\epsilon$.

\section{Slowly rotating regular black holes whose thin shell disappears in the no-rotation limit} 
Finally, let us consider the case where the thin shell inside the inner horizon disappears in the no-rotation limit, i.e., the case of $\sigma_0 =0$ and $p_0 = 0$.
In other words, we consider the case where although the de Sitter spacetime and the black hole spacetime are matched at some radius, $R$, 
there is no delta-function like matter distribution  between them in the unperturbed regular black hole solution. Since these solutions are 
characterized by $\sigma_0 =0$, some parts of the results are already given in Sec.~III. The physical quantities associated with the $\sigma_0=0$ cases 
correspond to those shown on the left vertical axis of all the figures given in Sec.~III. However, we reanalyze this case in this section because 
the unperturbed solutions and some perturbation quantities are given in simple analytic forms. As discussed later, the condition for determining 
the spherically symmetric perturbation differs from that supposed in Sec.~III, because the condition of $\sigma_0=0$ is exactly assumed in this section.  

The spherically symmetric solution for this case is investigated in detail by Lemos and Zanchin~\cite{lemos}. As argued in Ref.~\cite{lemos}, the charge, $Q$, and 
gravitational mass, $M$, of the spherically symmetric regular black hole may be, in terms of the matching radius, $R$, and the de Sitter horizon radius, $L$, given by  
\be
Q = \f {\sqrt{3} R^2} {L}\,, \quad M = \f {2 R^3} {L^2}\,.
\label{lemo1}
\ee
The regular black hole solution whose matching radius, $R$, satisfies $R<r_-$ and $R<L$ is allowed for the range of $R$, given by 
\be
\sqrt{3}/2 < R/L < 1\,,
\label{lemo2}
\ee
as argued in Refs.~\cite{lemos,uchi}.
Thus, regular black hole solutions considered by Lemos and Zanchin may be specified if values of $R/L$ satisfying Eq.~(\ref{lemo2}) are given. 

Using Eqs.~(\ref{lemo1}) and (\ref{lemo2}) for the unperturbed quantities of the slowly rotating regular black hole, we may evaluate the rotational 
effects on the regular black hole solution. The dipole perturbation quantities are then given by 
\beq
 \o &=& \f {a} {L^2} \,, \label{lemo3} \\ 
 \Omega &=& 0 \,, \label{lemo4} \\
 C_2 &=&\f {\sqrt{3}\, (R/L)^2} { R/L-X } \,,\label{lemo5}
\eeq
where $X=\mbox{arctanh} (R/L)$ as introduced before.
The matter rotation angular velocity, $\Omega$, vanishes at the $\epsilon$-order approximation because there is no matter field at the surface of 
the de Sitter sphere inside the inner horizon for the unperturbed solution. 

For the quadrupole perturbations, we obtain 
\be
\xi_2 = -\f {4 R^5f _{dS}} {3 L^4}\,, 
\label{lemo6}
\ee
\begin{align}
\zeta_2 & = \f {2 R^7} {L^7 (R- L X)^2} \left (- L R^3 (9L^4 -3 L^2 R^2 + 10 R^4) \right . \nn \\
&+ R^2 (27 L^6 - 6 L^4 R^2 + L^2 R^4 + 10 R^6) X \nn \\
& - L R (2 7 L^6 - 3 L^4 R^2 - 27 L^2 R^4 + 11 R^6 ) X^2 \nn \\
& + \left . 9L^4f_{dS}^2 (L^2 +R^2) ^2 X^3 \right ) \nn \\
& \times (R (-3 L^4 + 8 L^2 R^2 + 4 R^4 )+ 3 L^3 (L^2 -3 R^2) X) ^{-1}\,, 
\label{lemo7}
\end{align}
\begin{align}
D_2 & = \left \{\f{8 R^9} {3L^7 (R -L X)^2}\Big ( L R^2 (L^2-19R^2) \right.\nn \\
& + R (7L^4+ 2 L^2 R^2+ 15 R^4)X - 4L^3 f_{dS} (2 L^2 - 3 R^2) X^2 \Big ) \nn \\
& - 8 R^4 (\zeta_2 -\xi_2) \bigg \} (-3 L^2 R + 5  R^3 + 3 L^3 f_{dS}^2 X) ^{-1}.
\label{lemo8}
\end{align}
By using the relations given in Eqs.~(\ref{lemo5}) through (\ref{lemo8}), we may obtain values of $\delta \sigma_2$ and $\delta p_2$ 
through the relations, given by 
\begin{align}
\delta \sigma_2 & = \f{1}{\pi\sqrt{f_{dS}}}\left \{ \f{2R^3f_{dS} }{3L^4} +\f{\zeta_2 + 2 \xi_2-m_2}{4R^2}  +\f{f_{dS} k_2^{\prime}}{4} \right \} \,, 
\label{lemo9}
\end{align}
\begin{align}
\delta p_2& = \f {R^3} { 2 \pi L^4} \left (  \f { L^2 (L^2 - 2 R^2) m_2} {4 R^5 \sqrt {f_{dS}}^3} - \f {L^4 \sqrt {f_{dS}} (h_2 ^{\prime} + k_2 ^{\prime})} {4 R^3}\right . \nn \\
&  -\f {(L^4-6 L^2 R^2 + 6 R^4) \zeta_2 + L^2 (2L^2 -3R^2) \xi _2 } {4 R^5 \sqrt { f_{dS}} ^3} \nn \\
& \left . - \f {2 L^2 - 3 R^2} {3 L^2 \sqrt {f_{dS}}} \right ) \,. 
\label{lemo10}
\end{align}
Although $S^a_b=0$ in the no-rotation limit, the matter field in general appears on the surface 
of the de Sitter sphere as the rotation parameter, $a$, increases from zero. More specifically speaking, the energy density perturbation $\delta \sigma$ 
and the pressure perturbation $\delta p$ basically appear at the $\epsilon^2$-order approximation in the present treatment, as shown 
in Eqs.~(\ref{lemo9}) and (\ref{lemo10}). 

\begin{figure}[t]
\includegraphics {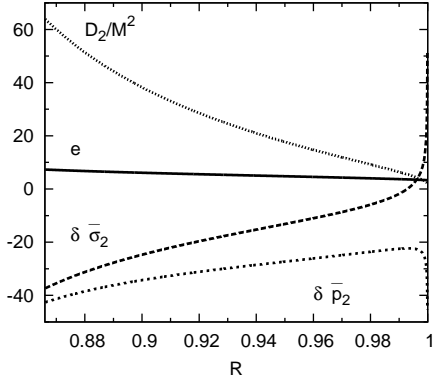}
\caption{The dimensionless integration constant, $D_2/M^2$, the eccentricity of the thin shell, 
$e/\epsilon$, the dimensionless energy density perturbation, $\delta\bar{\sigma}_2\equiv\delta \sigma_2/(M/4\pi R^2)$, 
and the dimensionless pressure perturbation, 
$\delta\bar{p}_2\equiv\delta p_2/(M/4\pi R^2)$, given as functions of the unperturbed matching radius, $R$.}
\label{noshell-l2}
\end{figure}
\begin{figure}[t]
\includegraphics {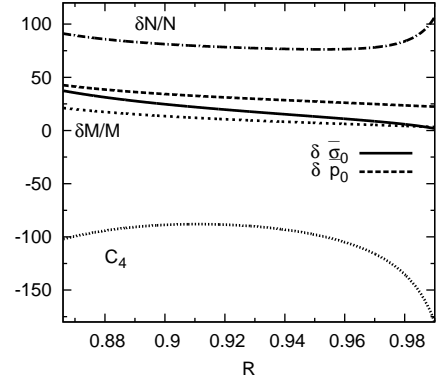}
\caption{Same as Fig. \ref{noshell-l2},  but for the relative change in the gravitational mass, $\delta M/M$, 
the dimensionless energy density perturbation, $\delta\bar{\sigma}_0\equiv\delta\sigma_0/(M/4\pi R^2)$, 
the dimensionless pressure perturbation, $\delta\bar{p}_0\equiv\delta p_0/(M/4\pi R^2)$, 
and the integral constant $C_4$.}
\label{noshell-l0}
\end{figure}
\begin{figure}[t]
\includegraphics {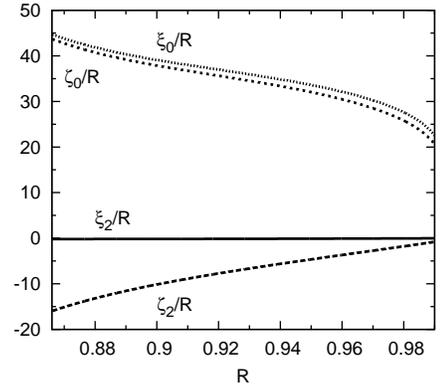}
\caption{Same as Fig. \ref{noshell-l2}, but for the dimensionless radial displacement of the matching radius, $\xi_0/R$, $\zeta_0/R$, $\xi_2/R$, and $\zeta_2/R$.}
\label{noshell-xz}
\end{figure}

Values of the quadrupole perturbation quantities, the dimensionless integration constant, $D_2/M^2$, the eccentricity of the thin shell, 
$e/\epsilon$, the dimensionless energy density perturbation, $\delta \sigma_2/(M/4\pi R^2)$, and the dimensionless pressure perturbation, 
$\delta p_2/(M/4\pi R^2)$, are given as functions of the unperturbed matching radius, $R$, in Fig.~\ref{noshell-l2}. 
All the physical quantities are rescaled by using the length scale of the de Sitter horizon $L$.
In Fig.~\ref{noshell-l2},  we observe that values of $\delta \sigma_2$ become positive for unperturbed solutions  having $0.995 \lesssim R/L < 1$ 
but values of $\delta p_2$ are always negative. This kind of peculiar behavior is already found in the cases of the unperturbed solutions with 
$\sigma_0 \neq 0$ as discussed in Sec.~III.  Thus, we  henceforth focus on the unperturbed solutions with $\sqrt{3}/2 < R/L  \lesssim 0.995$ 
in order to avoid problems related to $\delta\sigma_2=0$. 

When the spherically symmetric perturbations are considered, we confront another problem with $\delta\sigma$ in this case because $\sigma_0=0$. 
As shown in Fig.~\ref{noshell-l2}, $\delta\sigma_2$'s basically 
take negative values. Thus, it is possible for the total energy density to have a negative value somewhere regardless of values of $\epsilon$ 
if values of $\delta\sigma_0$ are smaller than some critical values. Since the appearance of negative energy density is unphysical, we do not 
consider the solutions with $\sigma <0$ in this study.  
To focus on positive energy density models only, in this study, we assume that $\delta \sigma_0 = |\delta \sigma_2|$ if $\delta \sigma_2<0$ 
and $\delta \sigma_0 = \delta \sigma _2/2$ if $\delta \sigma_2 >0$. Thanks to this assumption, 
$\delta\sigma=\delta \sigma_0 + \delta \sigma_2 P_2(\cos\theta)>0$ is always satisfied. 
The numerical results of the spherically symmetric perturbation quantities, the relative change in the gravitational mass, $\delta M/M$, 
the dimensionless energy density perturbation, $\delta\sigma_0/(M/4\pi R^2)$, the dimensionless pressure perturbation, $\delta p_0/(M/4\pi R^2)$, 
and the integral constant $C_4$ are shown in Fig.~\ref{noshell-l0}.
As mentioned before, we only show the results for the unperturbed solutions with $0.866  < R/L  < 0.99$ since values of $\delta \sigma_2$ become  positive
for $0.99  \lesssim R/L$. In this figure, we see that all the quantities given in the figure take positive values except for the case of $C_4$. 
Values of $C_4$ diverge in the limit of $R/L \to 1$ due to the terms proportional to $1/f_{dS}$, even though the results are not given in Fig.~\ref{noshell-l0}. 
The dimensionless radial displacement of the thin shell, $\xi_0/R$, $\zeta_0/R$, $\xi_2/R$, and $\zeta_2/R$, is given as functions of the matching radius 
of the unperturbed solution, $R$, in Fig. \ref{noshell-xz}.
\section{Conclusion}
We have obtained rotating solutions of regular black holes which are constructed of de Sitter spacetime with the axisymmetric stationary perturbation within 
the timelike charged thin shell and the Kerr-Newman geometry with sufficiently small rotation outside the shell. 
To treat the slowly rotating thin shell, we basically employ the method developed by de la Cruz and Israel~\cite{dlc2}. 
The thin shell is assumed to be composed of a dust in the zero-rotation limit and located inside the inner horizon of the black hole solution.
We expand the perturbation in powers of the rotation parameter of the Kerr-Newman metric, $a$, up to the second order. 
It is found that the thin shell in general has anisotropic pressure at the second-order approximation of the rotation parameter--i.e., 
the slowly rotating thin shell may not be in general composed of a dust with the present treatment.  
We may, however, set the shell to be composed of a perfect fluid with isotropic pressure by using the degree of freedom appearing in the physically acceptable 
matching of the two distinct spacetimes. 

By assuming the thin shell matter to be described by the perfect fluid, we investigate rotational effects on the regular black holes in detail. 
The dipole and quadrupole perturbations are uniquely determined by physical quantities of unperturbed spherically symmetric regular black holes and 
the rotation parameter, $a/M$. To obtain the spherically symmetric perturbation, we need to specify what equilibrium 
sequences of slowly rotating regular black holes are considered. In this study, 
we assume the following two cases:  (i) Values of the spherically symmetric energy density perturbation are specified in 
some appropriate way. (ii) The total particle number of the thin shell is conserved as the rotation parameter increases from zero. 
Under the assumption (ii), we numerically evaluate the perturbation quantities due to rotation for several equilibrium sequences of 
the unperturbed solutions of the spherically symmetric regular black hole characterized by a fixed gravitational mass, $M$, considered in Ref.~\cite{uchi}.
For the case of the equilibrium sequence with $M=1.9$, values of $\delta \sigma_2$ become positive for the unperturbed solutions with 
$4 \pi R^2 \sigma_0 \gtrsim 0.007$, while values of $\delta p_2$ are always negative. This is an unusual situation, because it implies that 
$dp/d\sigma <0$.  
The removal of this peculiar property with the matter field is in principle difficult with the present treatment, because the spacetime metric is a prior given 
then the matter properties read out through Einstein equations. 
Fortunately, this peculiar behavior is not found for the other equilibrium sequences considered in this study. 
The present analysis is also applied to the case where the slowly rotating thin shell disappears or the stress energy tensor of the thin shell 
vanishes in the zero-rotation limit. This case corresponds to choosing 
the spherically symmetric regular black hole solution considered in Ref.~\cite{lemos} as an unperturbed solution. The assumption (i) is used in this case 
in order to avoid the situation where a thin shell having negative total energy density is obtained.

The present results show that the regular black hole considered in Refs.~\cite{lemos,uchi} can rotate within an accuracy up to the second 
order of the spin parameter of the black hole. As mentioned in Sec.~III, the coefficients of the second-order perturbations in $V^-$ 
appear to be slightly larger than those of standard rotating objects like rotating stars (see, e.g., Refs.~\cite{hartle, chandra}). 
This fact might give a slightly stronger restriction for  suitable values of the dimensionless spin parameter $\epsilon$. 
Therefore, it is interesting and important to see whether the present $\epsilon^2$-order approximation results can be generalized to include 
higher-order rotational effects. Studies on rotating regular black holes with nonperturbative approaches are also interesting. In this study, 
we focus on the case of the perfect-fluid thin shell. However, the thin shell with anisotropic pressure might be more reasonable than 
the case of the perfect fluid, because the elemental field, like a scalar field naturally gives the stress energy tensor with the anisotropic pressure. 
Thus, it is not unimportant to investigate in detail the case of the non-perfect-fluid thin shell. 
These investigations remain as future work. 

\section*{Acknowledgements}
N.U. acknowledges financial support provided under the European Union's FP7 ERC Starting Grant 
``The dynamics of black holes: testing the limits of Einstein's theory,'' Grant Agreement No. DyBHo--256667. 
S.Y. thanks Jos{\'e} Lemos and Vitor Cardoso for their kind hospitality at Institute Superior T{\'e}cnico, 
where some parts of this work were done. 
This work was supported in part by a Grant-in-Aid for Scientific Research from JSPS (Grant No. 24540245). 
\appendix
\section{Metric and Electromagnetic fields in $V^-$}
In this appendix, we briefly describe the metric and electromagnetic fields assumed in $V^-$. Following the standard 
prescription given in Ref.~\cite{hartle} (see also Refs.~\cite{thorne,chandra,friedman}), we consider stationary and 
axisymmetric perturbations around the de Sitter solution given in the static coordinates. The metric and 
electromagnetic field perturbations are given as solutions of  
the perturbed vacuum Einstein-Maxwell equations including the cosmological constant $\Lambda$, 
\be
\Delta \left[ {G_{\mu}}^\nu +\Lambda \, {\delta_\mu}^\nu \right] 
= \Delta \left[ 2 \left( F_{\mu\alpha}F^{\nu\alpha} -{1\over 4}\,{\delta_\mu}^\nu F_{\alpha\beta}F^{\alpha\beta} \right) \right] \,, 
\label{ein}
\ee
\be
\Delta \left[ \nabla_\alpha F^{\mu\alpha} \right]=0 \,, 
\label{max}
\ee
where ${G_{\mu}}^\nu$ and $\nabla_\alpha$ mean the Einstein tensor and covariant derivative with respect to 
the perturbed metric, respectively, and $\Delta q$ denotes the perturbation of the quantity $q$.  In this study, as mentioned 
in the main text, we assume the equatorial symmetry and the time-azimuth reflection symmetry for the spacetime and 
electromagnetic field considered, which are reflected by the symmetry of the Kerr-Newman solution assumed as $V^+$. 
Taking account of the  rotational effects up to the second order of $\epsilon \equiv a/M$, we may therefore assume the metric and 
vector potential as follows. 

\noindent
The line element: 
\be
\begin{split}
& ds^2= \\ 
& -f_{dS}(r)\left\{ 1+2 \epsilon^2 \left( h_0(r)+h_2(r)P_2(\cos\theta) \right) \right\}  dt^2 \\
&\displaystyle{ +f_{dS}^{-1}(r)\left\{ 1+ \f{2  \epsilon^2 (m_0(r)+m_2(r)P_2(\cos\theta)) }{r f_{dS}(r)} \right\} dr^2} \\
& + r^2\left( 1+2\epsilon^2 k_2(r)P_2(\cos\theta) \right)  \\ 
&\quad\quad  \times \left\{ d\theta^2 +\sin^2 \theta \left( d\phi - \o(r) dt \right)^2 \right\}+O(\epsilon^3) \, ,
\end{split}
\ee
where $f_{dS}(r)\equiv1-r^2/L^2$, $\omega(r)=O(\epsilon)$, and $P_l$ denotes the Legendre polynomial  of degree $l$. 
Here, we employ the gauge condition for the metric perturbation the same as that of Ref.~\cite{hartle}. For the regular black 
hole considered in this study, we assume that $0 \le r <L$. 

\noindent
The vector potential: 
\be
\begin{split}
\tilde{A}&= \left\{ -{Q\over R}+\epsilon^2 \left( B_0(r)+B_2 (r)P_2(\cos \theta) \right) \right\} dt \\ 
& \quad\quad  + a A_3(r) \sin ^2 \theta \, d\phi +O(\epsilon^3) \, ,
\end{split}
\ee
where $R$ and $Q$ denote the radius and charge of the thin shell in the limit of $\epsilon\rightarrow 0$, respectively. 

The $\phi$ component of the Maxwell equation, $\Delta \left[ \nabla_\alpha F^{\phi\alpha} \right]=0$, leads the $\epsilon$-order 
equation, 
\be
 A_3 ^{\prime \prime} +{f_{dS}^{\prime}\over f_{dS}}\, A_3 ^{\prime}  -\f{2 }{r^2 f_{dS}}\, A_3 =0\,, 
\ee
where the prime ($^\prime$) means the derivative with respect to $r$. The general solution of this equation is given by
\be
 A_3 = \f {C_1}{r} +C_2 \, \f {r- L\, \mbox{arctanh} (r/L)}{r} \,, 
\ee
where  $C_1$ and $C_2$ are integral constants. From the regularity of the electromagnetic fields at $r=0$, we need to impose 
$C_1 =0$. We then have 
\be
 A_3 =C_2 \, \f {r- L\, x }{r} \,, 
\ee
where 
\be
x\equiv \mbox{arctanh} ({r\over L}) \,.
\ee
The other $\epsilon$-order equation is obtained from $\Delta [{G^t}_\phi] =0$, given by 
\be
\left(r^4 {\o}^{\prime} \right)^{\prime}=0 \,. 
\ee
From the regularity at $r=0$, thus, we have
 \be
 {\o}={\rm constant} \,.  
\ee

Since the regular solutions within an accuracy up to the order of $\epsilon$ have been obtained, we may move on to considerations of the next-order solutions. 
The $t$ component of the Maxwell equation, $\Delta \left[ \nabla_\alpha F^{t \alpha} \right]=0$, leads the $\epsilon^2$-order 
equations, given by 
\be
\begin{split}
& B_2 ^{\prime \prime}  + \f{2 B_2 ^{\prime} }{r} - \f{ 6 B_2}{r^2 f_{dS}} = \\
&\quad 4 M^2 \o_1  C_2 \left\{ 
\f{L \, \mbox{arctanh} (r/L)}{r^3 f_{dS}} - \f{ 3-2(r/L)^2 }{ 3\, r^2 f_{dS}^2 } 
\right\} \,, 
\end{split}
\ee
\be
B_0^{\prime \prime}  +\f{2 B_0 ^{\prime}}{r} = \f{4\, M^2 \o_1  C_2}{3 L^2 f_{dS}^2} \,, 
\ee
where $\o \equiv a \o_1$ with $\o_1$ being a constant of $O(1)$. 
The general solutions of these equations are given by 
\begin{align}
B_2 & = -\f{a_0 L^2 f_{dS}}{ r^3}+ a_1\f {-3 L^2 r + 2 r^3 + 3 L^3 f_{dS} x}{2 r^3}  \nn \\
&  + \f{M^2 \o_1 L C_2}{6 r^3} \left\{ -L^2 f_{dS} \log y -4 L^2 x +6 L r \right\} \,, \\
B_0 &=a_2 -\f{a_3}{r} + \f{2 \, L M^2 \, \o_1  C_2 \, x}{3 r} \,,
\end{align}
where
\be
 y \equiv \f{L+r}{L-r} \,. 
\ee
Here, $a_0$, $a_1$, $a_2$ and $a_3$ are integral constants.
Near the origin, $r \sim 0$, these solutions become  
\begin{align}
B_2 & = -\f{L^2 a_0}{r^3}  +\f{a_0}{r} +O(r^2) \, , \\
B_0 & = - \f{a_3}{r}  + \left (a_2 + \f{2 \o_1 C_2 }{3 M}  \right ) +O(r^2) \,. 
\end{align}
Therefore, if we choose 
\begin{align}
 a_0 =0 \,, \quad  a_3 =0\,,
\end{align}
we have the regular solutions, given by  
\be
\begin{split}
B_2 & = \f{M^2 C_2 L\o_1}{3 r^3}\left\{ 3 L r +(r^2-3 L^2) x \right\}  \\
&+ \f{a_1}{2  r^3} \left ( 2r^3-3 L^2 r+  3  L^3 f_{dS} x\right ) \,,\\
 B_0 & = a_2 + \f{2 \, M^2 L \, \o_1 \, x }{3 r} \,,
\end{split}
\ee
where we have used the relation $2 x = \log y$. Note that this identity will be frequently used in the following algebraic manipulation. 

The other $\epsilon^2$-order equations are obtained from the perturbed Einstein equations. As can be seen from Eq.~(\ref{ein}), 
the right-hand sides of the Einstein equations are composed of quadratic forms in $A_3(r)$ and $A_3^{\prime}(r)$. The independent 
set of the master equations for $h_0(r)$, $m_0(r)$, $h_2(r)$, $m_2(r)$, and $k_2(r)$ are summarized as follows: 
\begin{align}
m_0^\prime =  \frac{M^2}{3 r^2}  \left\{ 2 (A_3)^2+f_{dS}  r^2 (A_3 ')^2\right\} \,, 
\label{m_0_eq}
\end{align}
\begin{align}
& h_0^\prime =  \frac{1-3 (r/L)^2}{ r^2 f_{dS}^2} \, m_0 \nn \\
&\quad\quad +\frac{M^2}{3 r^3}\, \left\{  r^2 (A_3')^2-2 f_{dS}^{-1} (A_3)^2\right\}   \,, 
\label{h_0_eq}
\end{align}
\begin{align}
m_2 = - r f_{dS} \left\{ h_2-{2\over 3} M^2 f_{dS}  (A_3')^2\right\}  \,, 
\label{m_2_eq}
\end{align}
\begin{align}
k_2^\prime +h_2^\prime=  \frac{1-2 (r/L)^2}{r^2 f_{dS}^2} \, m_2
+\frac{h_2}{r\,f_{dS}}+\frac{4 M^2 A_3 A_3' }{3 r^2} \,, 
\label{k_2_eq}
\end{align}
\begin{align}
&h_2^\prime=  \frac{2 (r/L)^2-1}{f_{dS}}\,  k_2^\prime+\frac{1-3 (r/L)^2}{r^2 f_{dS}^2}\, m_2 \nn \\
&\quad\quad +\frac{2}{r f_{dS}}\, k_2 + \frac{3}{r f_{dS}}\, h_2  \nn \\
&\quad\quad -\frac{M^2}{3 r^3}\,  \left\{r^2 (A_3')^2+4 f_{dS}^{-1} (A_3)^2\right\}  \,, 
\label{h_2_eq}
\end{align}
\begin{align}
&k_2^{\prime \prime}= \frac{1}{r^2 f_{dS}} \, m_2^\prime - \frac{3-4 (r/L)^2}{r f_{dS}} \, k_2^\prime  \nn \\
&\quad\quad +\frac{3}{r^3 f_{dS}^2}\, m_2 + \frac{2}{r^2 f_{dS} }\, k_2 \nn \\
&\quad\quad +\frac{M^2 }{3 r^4}\left\{ r^2 (A_3')^2-4 f_{dS}^{-1} (A_3)^2\right\}  \,.
\label{dk_2_eq}
\end{align}
From Eqs. (\ref{m_0_eq}) and (\ref{h_0_eq}), we obtain their general solutions, given by 
\begin{align}
& m_0 = C_3 +\f{M^2 (C_2) ^2}{6 L r^3}\left\{ 4 L^2 r x-2 L^3 f_{dS}\, x^2 \right. \nn\\
& \quad\quad\quad\quad\quad\quad\quad\quad\quad - \left.r^2 (2 L + r \log y) \right\} \,, \\
& h_0= C_4+ \f{M^2 (C_2) ^2}{6 L^3 r^4f_{dS} }\left\{ -2 r(L^4-2L^2 r^2 +3r^4) x\right . \nn \\
& \quad\quad\quad\quad\quad\quad\quad\quad + L f_{dS} (L^4 -3  r^4 )x ^2  \nn \\ 
& \quad\quad\quad\quad\quad\quad\quad\quad  \left. + r^2 L ( L^2 + 3r^2 +L r \log y) \right\} \,, 
\end{align} 
where $C_3$ and $C_4$ are integral constants. These functions can be written near the origin by 
\begin{align}
m_0 &= C_3 + O(r^3) \, ,\\
h_0 &= \f{M^2 (C_2)^2}{3 L^2 }+ C_4 + O(r^3) \,.
\end{align}
Regularity at $r=0$ requires that $m_0 \to 0$ as $r \to 0$; we then need to set  
\be
C_3 =0\,. 
\ee
We then obtain the regular solutions:  
\begin{align}
& m_0 =- \f{M^2 (C_2) ^2}{3 r}\left (\f{r^2 -2 L^2}{L r} x+\f{L^2 f_{dS}x^2}{r^2} +1  \right) \,, \\
& h_0= C_4+ \f{M^2 (C_2) ^2}{3 L^2 r^2f_{dS} }\left\{ - \f{L^4 -3 L r^2 +3 r^4}{L r} x\right . \nn \\
& \quad\quad\quad\quad\quad\quad \left. +\f{(L^4-3r^4)f_{dS}x^2 }{2 r^2} + \f{L^2 +3r^2}{2} \right\} \,.
\end{align} 
By using Eqs.~(\ref{m_2_eq})--(\ref{dk_2_eq}), we may obtain the second-order ordinary differential equation for $h_2$, 
given by 
\be
\begin{split}
&   L^2 f_{dS} ^2 h_2^{\prime \prime} +2 \f{(L^2 - 2 r^2 )}{r}  f_{dS}  h_2^{\prime } -  \f{2}{r^2 L^2}(3 L^4-r^4)  h_2  \\
& -\f{2(C_2) ^2 M^2}{3r^6 f_{dS}}   \left\{ r^2( 3L^4  -11 L^2 r^2 + 4r ^4 )+3 L^6  f_{dS} ^3 x^2 \right . \\
& \left . -2L ^3 r (3 L^2 -7 r^2)  f_{dS} x  \right\}= 0 \, .
\end{split}
\ee
The general solution of this equation is given by 
\be
\begin{split}
h_2 &=- D_1 \f {f_{dS}} {r^3}  +D_2\left \{ \f{1}{ 8 f_{dS} }\left (\f{5}{L^2} -\f{3}{r^2} \right ) +\f{3Lf_{dS}}{8r^3}\,x \right \} \\
&+\frac{(C_2)^2 M^2 }{24 r^4 L^3 f_{dS}}  \left[ -16 L^2 r^3 x +8 L^5 f_{dS}^2 x^2 \right. \\
& \left. +r \left\{  2 L r \left(L^2+5 r^2\right)-5 L^4 f_{dS}^2 \log{y} \right\}  \right] \,,
\end{split}
\ee
where $D_1$ and $D_2$ are integral constants.
If $r \sim 0$, this solution becomes 
\be
\begin{split}
h_2 &= -\f{D_1}{r^3 } + \f{ L^2 D_1}{r} + O(r^2) \, . 
\end{split}
\ee
To have a regular solution near the center, therefore, we have to impose 
\be
D_1 =0\,. 
\ee
Then, the regular solution $h_2$ at the origin is obtained. From this regular solution $h_2$, 
the regular solutions $m_2$ and $k_2$, are given by
\begin{align}
 m_2  &= -f_{dS}  r h_2 + \f{2M^2 (C_2) ^2} {3 r^3 } \left (r -L f_{dS}x \right )^2 \, , \\
 k_2  &= \f{r^3} {2L^2} h_2^{\prime } +\left(\f {r^2 } {2 L^2} -\f {1 } {f_{dS}} \right )h_2 \nn \\
 & + \f{M^2 (C_2) ^2 } {6 r^2 L^4 f_{dS} } \left\{   \f {L^4+ 3 L^2 r^2 -2 r^4} { f_{dS}}  \right . \nn \\
 &  \quad\quad\quad\quad\quad -\f {2L} {r} (L^4 + 5 L^2 r^2 -2 r^4)x  \nn \\
 &  \quad\quad\quad\quad\quad  \left. +  \f{f_{dS} L^2} {r^2} (L^4 + 7 L^2 r^2 +2 r^4) x^2 \right \} \,.
\end{align}
We finally obtain the set of regular solutions, given by 
\begin{align}
h_2 &= D_2\left \{ \f{1}{ 8 f_{dS}}\left (\f{5}{L^2} -\f{3}{r^2} \right ) +\f{3L f_{dS}}{8r^3}\, x \right \} \nn\\
& + M^2 (C_2)^2 \left\{ \f{L^2 f_{dS}x^2 }{3 r^4}  \right . \nn \\
& \ \ \left . + \f{Lr (L^2 +5r^2) + ( - 5 L^4 +2 L r^2 -5r^4) x}{12 L^3 r^3 f_{dS} }\right \} \,,
\end{align}
\begin{align}
 m_2 & = \f {D_2 } {8 L^2 r^2} \left\{ r (3 L^2 -5r^2) - 3L^3 xf_{dS} ^2  \right\}  \nn \\
 &+ M^2 (C_2)^2   \left ( \f{5r^4+14 L^2 r^2 -11 L^4}{12 L^3 r^2} x  \right . \nn \\
& \quad\quad\quad\quad \left . +\f{L^2 f_{dS}^2 x^2}{3r^3}+\f{7L^2 -5r^2}{12rL^2}\right ) \,, 
\end{align}
\begin{align}
 k_2 &= \f {   D_2} {8 r^3 L^2 }\left\{ r(3 L^2 +4 r^2) -3 L(L^2 + r^2 )x\right\}  \nn \\
& +M^2 (C_2) ^2 \left(\f{L^2 -7r^2}{12 L r^3}x  \right. \nn\\ 
& \quad\quad\quad\quad \left .-\f{L^4 -4 L^2 r^2 -3 r^4}{6 L^2 r^4}x^2+\f{1}{12r^2} \right)\,.
\end{align} 

\section{The $\epsilon^2$-order coefficients of the extrinsic curvature}

The $\epsilon^2$-order coefficients of the extrinsic curvature  associated with $\Sigma$ are summarized in this appendix. 
Nonzero coefficients of $^{(2)}{K^\pm}^a_b$ are given for the sake of saving space, 
where $^{(2)}{K^\pm}^a_b$ is defined by 
\be
^{(2)} {K^\pm}^a_b \equiv {1\over 2 !} \lim_{\epsilon \to 0}{ \partial^2\over \partial \epsilon^2} {K^\pm}^a_b \,. 
\ee
\begin{align}
^{(2)}{K^+}^T_T &=  \f{1}{ R^3 \sqrt{f_{RN}}^3}\biggl[  \delta M(M-R)  \nn \\
& + \f{M^2}{6R^4} \left\{ Q^4-Q^2 R (M+2 R)-2 M R^2 (M-2 R) \right\}  \nn \\
 & \left.- \f{\zeta_0}{R^3} \left\{ Q^2 (3 R^2 -6M R + 2 Q^2)+M R^2 (3 M-2R)\right\} \right ]\nn \\
 & + \f{P_2 }{R^5\sqrt{f_{RN}}} \left[ \f{M^2(7M R-5Q^2)}{3} \right. \nn\\
 &\quad\quad\quad\quad -\f{\zeta_2}{R f_{RN}} \left\{ 3Q^2 (R^2 - 2 M R ) \right. \nn \\
 & \quad\quad\quad\quad\quad\quad \left.+ 2 Q^4 + 3M ^2 R^2 -2M R^3\right\} \biggr] \,,
\end{align}
\begin{align}
^{(2)} {K^+}^{\Theta}_{\Theta} &+^{(2)}{K^+}^{\Phi}_{\Phi} \nn \\
& = \f {1} { R^4 \sqrt {f_{RN}}} \left \{ \f {M^2 \left(2 R^2 - 4 M R + Q^2\right) } {3 R}  \right . \nn \\ 
& \quad\quad +2 \delta M R^2 + 2 \left(2 Q^2 - 3 M R + R^2 \right) \zeta_0 \bigg\} \nn \\
& + \f {2P_2} {R^4 \sqrt {f_{RN}} } \left\{  \f {M^2 \left(4 Q^2 - 7 M R + 2 R^2\right)} {3 R}\right. \nn \\
& \quad\quad +  \left(2 Q^2 - 3 M R + R^2 \right)\zeta_2 \bigg\} \, , 
\end{align}
\begin{align}
^{(2)} {K^+}^{\Theta}_{\Theta} &- ^{(2)}{K^+}^{\Phi}_{\Phi} \nn \\
 & = - \f {M^2 (R^2 + RM - Q^2)} {R^5 \sqrt {f_{RN}}} \sin^2 \Theta \,, 
 \end{align} 
 \begin{align}
 ^{(2)}{K^+} &= \f{1}{6R^7 \sqrt{f_{RN}}^3 } \left \{ M^2(3 Q^4 + Q^2 R (4 R -13 M) \right .\nn \\
  & +2R^2(7M^2-6MR +2 R^2) ) +6R (2Q^4  \nn \\
  &+ Q^2 R (3R-8M) +R^2 (9M^2 - 8M R +2 R^2) ) \zeta_0   \nn \\
  & + 6 \delta M R^3 (2Q - 3M R+R^2)  +2 \left (M^2 (3 Q^4  \right .\nn \\
  & + Q^2 R (7R-13 M)  +R^2 (14 M^2 -15 M R +4 R^2)) \nn \\
  &+3R (2 Q^4 -Q^2 R (8 M+3 R)   \nn \\
  & \left. \left. +R^2 (9 M^2 + 4 M R -4 R^2) ) \zeta_2 \right )P_2  \right \} \,, 
\end{align} 
\begin{align}
^{(2)}{K^-}^{T}_{T} &= \f{1}{L^2 (\sqrt{f_{dS}})^3 } \left\{  \xi_0 -m_0 -L ^2 f_{dS}^2 \, h_0^{\prime}\right .\nn \\
 & \quad\quad\quad\quad \left. +\left( \xi_2 -m_2 -L^2 f_{dS}^2 \, h_2^{\prime} \right) P_2 \right\} \,, 
 \end{align} 
 \begin{align}
 ^{(2)} {K^-}^{\Theta}_{\Theta} & + ^{(2)}{K^-}^{\Phi}_{\Phi} \nn \\
 & = \f {2 (\xi_0 + m_0)} {R^2 \sqrt {f_{dS}}} + 2 P_2 \left ( \f {-2 \xi_2 + m_2} {R^2 \sqrt {f_{dS}}}  - \sqrt {f_{dS}} k_2 ^{\prime} \right ) \,, 
 \end{align} 
 \begin{align}
 ^{(2)} {K^-}^{\Theta}_{\Theta} &-^{(2)}{K^-}^{\Phi}_{\Phi} = \f {3 \xi_2} {R^2 \sqrt {f_{dS}}} \sin ^2 \Theta \, , 
  \end{align} 
 \begin{align}
^{(2)}K^- = 
&  \f{1}{L^2 R^2 (\sqrt{f_{dS}})^3} \Biggl[  (2L^2 -R^2) \, \xi_0+ (2L^2-3R^2)\, m_0   \nn \\
& -L^2 R^2 (f_{dS})^2 h_0^{\prime}+\biggl\{  - (4L^2 -5 R^2) \xi_2   \nn \\
&  +(2L^2 -3 R^2)\, m_2 - L^2 R^2 (f_{dS})^2 (h_2^{\prime} +2  k_2^{\prime})\biggr\}  P_2 \Biggr] \,.
\end{align} 

\section{The explicit expressions for  $\delta\sigma_0$ and  $\delta\sigma_2$}
%
\beq
\delta \sigma_0& =& \f {1} { 4 \pi R^2 } \left[ \f {1 } { \sqrt {f_{RN}} } \left\{ \f{ (R^2 -3 M R + 2 Q^2)\zeta _0} {R^2} + \delta M  \right. \right\} \nn \\
&& \left .- \f {\xi_0 + m_0} {\sqrt {f_{dS}}} \right]  +\frac{M^2 \left(Q^2-2 M R+R^4 \Omega_1 \right)^2}{12 \pi  R^7 \sqrt{f_{dS}}\, f_{RN}} \nn \\ 
&& +{M^2\over 24\pi R^7 \sqrt{f_{RN}}^3}\Biggl[ 2 R^2 \left(6 M^2-4 M R+R^2\right)  \nn \\
&&\quad\quad\quad\quad\quad+3 Q^2 R (R-4 M)+3 Q^4  \nn \\
&&\quad\quad\quad\quad\quad+ 4 R^4 \left(Q^2-M R\right) \Omega_1 \nn \\
&&\quad\quad\quad\quad\quad-2 R^6 \left(R (R-3 M)+2 Q^2\right) \Omega_1^2
\Biggr] \,, 
\eeq
\beq
\delta \sigma_2 &=&\f{1}{4 \pi R^2} \left \{ \f {2 \xi_2 -m_2} {\sqrt {f_{dS}}}  
+ \f {(R^2 - 3 M R + 2 Q^2) \zeta_2} {R^2 \sqrt {f_{RN}}}  \right \} \nn \\
&&+\f{\sqrt{f_{dS}}}{4\pi}k_2^{\prime} 
-\frac{M^2 \left(-2 M R+Q^2+R^4 \Omega_1 \right)^2}{12 \pi  R^7 \sqrt{f_{dS}}\,f_{RN}} \nn \\
&&+{M^2\over 12 \pi R^7 \sqrt{f_{RN}}^3}\Biggl[ 2 R^4 \left(M R-Q^2\right)\Omega_1 \nn \\
&&+R^6 \left\{ R (R-3 M)+2 Q^2\right\}\Omega_1^2 +6 Q^2 R (R-2 M) \nn \\
&&+R^2 \left(12 M^2-11 M R+2 R^2\right)+3 Q^4
\Biggr] \,,
\eeq
where $\delta\sigma =\delta\sigma_0+\delta\sigma_2 P_2$.


\begin{thebibliography}{99}
\bibitem{pen} R. Penrose, Phys. Rev. Lett. {\bf 14}, 57 (1965). 

\bibitem{haw} S. Hawking and R. Penrose, Proc. R. Soc. A {\bf 314}, 529 (1970).                                                                                                                                                                                                                                                                                                                                                                                                                                                                                                                                                                                                                                                                                                                                                                                                                                                                                                                                                                                                                                                                                                                                                                                                                      

\bibitem{he} S. Hawking and G. F. R. Ellis, {\it The Large Scale Structure of Space-time} (Cambridge University Press, Cambridge, England, 1973). 

\bibitem{bro} K. A. Bronnikov, H. Dehen and V. N. Melnikov, Gen. Relativ. Gravit. {\bf 39}, 973 (2007).

\bibitem{an} S. Ansoldi, arXiv:0802.0330 (2008).

\bibitem{bar} J. M. Bardeen, in {\it Proceedings of GR5} (URSS, Tbilisi, 1968). 

\bibitem{bro2} K. A. Bronnikov, Phys. Rev. Lett. {\bf 85}, 4641 (2000); Phys. Rev. D {\bf 63}, 044005 (2001).

 \bibitem{ab} E. Ay\'on-Beato and A. Garcia, Phys. Rev. Lett.  {\bf 80}, 5056 (1998) ;  Gen. Relativ. Gravit.  {\bf 31}, 629 (1999) ; Phys. Lett. B {\bf 464}, 25 (1999).

\bibitem{ab2} E. Ay\'on-Beato and A. Garcia, Phys. Lett. B  {\bf 493}, 149 (2000).

\bibitem{ab3} E. Ay\'on-Beato and A. Garcia, Gen. Relativ. Gravit. {\bf 37}, 635 (2005). 

\bibitem{more} C. Moreno and O. Sarbach, Phys. Rev. D {\bf 67}, 024028 (2003).


\bibitem{dym} I.  Dymnikova, Gen. Relativ. Gravit. {\bf 24}, 235 (1992). 


\bibitem{dym2} I. Dymnikova and E. Galaktionov, Classical Quantum Gravity {\bf 22}, 2331 (2005).

\bibitem{hay} S. A. Hayward, Phys. Rev. Lett. {\bf 96}, 031103 (2006). 

\bibitem{dym3} I. Dymnikova and E. Galaktionov, Classical Quantum Gravity {\bf 21}, 4417 (2004).

\bibitem{bambi} C. Bambi and L. Modesto, Phys. Lett. B {\bf 772}, 329 (2013).


\bibitem{nino} A. Flachi and J. P. S. Lemos, Phys. Rev. D {\bf 87}, 024034 (2013).

\bibitem{sa}  A. D.  Sakharov, Sov. Phys. JETP  {\bf 22}, 241 (1966). 

\bibitem{gl}  \'E. B.  Gliner, Sov. Phys. JETP  {\bf 22}, 378 (1966). 

\bibitem{mar}  M. A. Markov, Sov. Phys. JETP Lett.  {\bf 36}, 265 (1982). 

\bibitem{fmm} V. P. Frolov, M. A. Markov, and V. F. Mukhanov, Phys. Lett. B {\bf 216}, 272 (1989);  Phys. Rev. D {\bf 41}, 383 (1990).

\bibitem{bal} R. Balbinot and E. Poisson, Phys. Rev. D {\bf 41}, 395 (1990).

\bibitem{lake} K. Lake and T. Zannias, Phys. Lett. A {\bf 140}, 291 (1989).

\bibitem{lemos} J. P. S.  Lemos and V. T. Zanchin, Phys. Rev. D {\bf 83}, 124005 (2011). 

\bibitem{uchi} N. Uchikata, S. Yoshida and T. Futamase, Phys. Rev. D {\bf 86}, 084025 (2012).

\bibitem{dlc2} V. de la Cruz and W. Israel, Phys. Rev. {\bf 170}, 1187 (1968).

\bibitem{israel} W. Israel, Nuovo Cimento B {\bf 44}, 1 (1966).

\bibitem{ba} C. Barrab\'es and W. Israel, Phys. Rev. D {\bf 43}, 1129 (1991).

\bibitem{ku} K. Kucha\u{r}, Czech. J. Phys. B {\bf 18}, 435 (1968).

\bibitem{jackson} J. D. Jackson, {\it Classical Electrodynamics} (Wiley, New York, 1998). 

\bibitem{hartle} J. B. Hartle, Astrophys. J. {\bf 150}, 1005 (1967); J. B. Hartle and K. S. Thorne, Astrophys. J. {\bf 153}, 807 (1968).

\bibitem{chandra} S. Chandrasekhar and J. C. Miller, Mon. Not. R. Astron. Soc. {\bf 167}, 63 (1974).

\bibitem{thorne} K. S. Thorne, {\it General Relativity and Cosmology}, 
edited by R. K. Sachs (Academic Press, New York, 1971).

\bibitem{friedman} J. L. Friedman and N. Stergioulas, {\it Rotating Relativistic Stars} 
(Cambridge University Press, New York, 2013). 

\bibitem{rw} T. Regge and J. Wheeler, Phys. Rev. {\bf 108}, 1063 (1957).

\end{thebibliography}
\end{document}